\documentclass{amsart}

\usepackage{amssymb}
\usepackage{amsfonts}
\usepackage{amsmath}
\usepackage{graphicx}%
\usepackage{lastpage}
\usepackage[utf8]{inputenc}
\usepackage[legalpaper,bookmarks=true,colorlinks=true,linkcolor=blue,citecolor=blue]{hyperref}
\usepackage{fancyhdr}
\usepackage{color}
\usepackage[mathlines]{lineno}
\usepackage{lscape}
\usepackage{epsfig}
\usepackage{natbib}
\usepackage{float}

\newtheorem{theorem}{Theorem}
\theoremstyle{plain}

\newtheorem{corollary}{Corollary}

\newtheorem{remark}{Remark}

\newtheorem{assumption}{Assumption}

\numberwithin{equation}{section}

\begin{document}

\title{Divergence Measures Estimation and Its Asymptotic Normality Theory Using Wavelets Empirical Processes}

\author{$^{(1)}$  Amadou Diadi\'e BA}
\email{amadou-diadie.ba@edu.ugb.en}
\author{$^{(1,2,3)}$Gane Samb LO}
\email{gane-samb.lo@ugb.edu.sn}
\author{$^{(1)}$Diam BA}
\email{diamba79@gmail.com}

\begin{abstract}
In this paper we provide the asymptotic theory of the general of $\phi$-divergences measures, which includes the most common divergence
measures : Renyi and Tsallis families and the Kullback-Leibler measure. Instead of using the Parzen nonparametric estimators of the probability density functions
whose discrepancy is estimated, we use the wavelets approach and the geometry of Besov spaces. One-sided and two-sided statistical tests are derived as well as symmetrized estimators.
Almost sure rates of convergence and asymptotic normality theorem are obtained in the general case, and next particularized for the Renyi and Tsallis families and for the Kullback-Leibler measure as well. The applicability of the results to usual distribution functions is addressed.

\bigskip
\bigskip \noindent $^{(1)}$ \textit{Amadou Diadi\'e Ba (amadou-diadie.ba@edu.ugb.en), Diam Ba (diamba79@gmail.com)}. LERSTAD, Gaston Berger University, Saint-Louis,
S\'en\'egal.\newline
\noindent $^{(2)}$ LSTA, Pierre and Marie Curie University, Paris VI, France.\newline
\noindent $^{(3)}$  AUST - African University of Sciences and Technology,
Abuja, Nigeria\newline

\noindent \textit{Corresponding author}. Gane Samb Lo. Email :
gane-samb.lo@edu.ugb.sn, gslo@aust.edu.ng, ganesamblo@ganesamblo.net\\
Permanent address : 1178 Evanston Dr NW T3P 0J9,
Calgary, Alberta, Canada.
\end{abstract}

\maketitle

\section{Introduction} \label{sec_intro} \label{sec1}

\subsection{General Introduction} \label{subsec_intro_generale}$ $\\

\noindent In this paper, we deal with divergence measures estimation using essentially wavelets density function estimation. Let $\mathcal{P}$ be a class of probability measures on $\mathbb{R}^{d}$, $d\geq 1$, a divergence measure on $\mathcal{P}$\ is a function 
\begin{equation}
\begin{tabular}{cccl}
$\mathcal{D}:$ & $\mathcal{P}^{2}$ & $\longrightarrow $ & $\overline{\mathbb{%
R}}$ \\ 
& $(\mathbb{Q},\mathbb{L})$ & $\longmapsto $ & $\mathcal{D}(\mathbb{Q},%
\mathbb{L})$
\end{tabular}
\label{divApp}
\end{equation}

\noindent such that $\mathcal{D}(\mathbb{Q},\mathbb{Q})=0$ for any $\mathbb{Q}$ such that $(\mathbb{Q},\mathbb{Q})$ in the domain of application of 
$\mathcal{D}$.\\

\noindent The function $\mathcal{D}$ is not necessarily an application. And if it is, it is not always symmetrical and it does neither have to be a metric. In case of lack of symmetry, the following more general notation is more appropriate :

\begin{equation}
\begin{tabular}{cccl}
$\mathcal{D}:$ & $\mathcal{P}_1 \times \mathcal{P}_2$ & $\longrightarrow $ & $\overline{\mathbb{%
R}}$ \\ 
& $(\mathbb{Q},\mathbb{L})$ & $\longmapsto $ & $\mathcal{D}(\mathbb{Q},%
\mathbb{L})$,
\end{tabular}
\label{divAppG}
\end{equation}

\noindent where $\mathcal{P}_1$ and $\mathcal{P}_2$ are two families of probability measures on 
$\mathbb{R}^d$, not necessarily the same. To better explain our concern, let us introduce some of
the most celebrated divergence measures.\\

\noindent A great number of them are based on probability density functions (\textit{pdf}). So let us suppose that any $\mathbb{Q}\in 
\mathcal{P}$ admits a \textit{pdf} $f_{\mathbb{Q}}$ with respect to a $\sigma $-finite measure $\nu $ on $(\mathbb{R}^{d},\mathcal{B}(\mathbb{R}^{d}))$, which is  usually the Lebesgue measure $\lambda_k$ (with $\lambda_1=\lambda$) or a counting measure on $\mathbb{R}^{d}$.\\

\noindent We may present the following divergence measures.\\

\noindent (1) The $L_{2}^{2}$-divergence measure :

\begin{equation}
\mathcal{D}_{L_{2}}(\mathbb{Q},\mathbb{L)=}\int_{\mathbb{R}^{d}}(f_{\mathbb{Q%
}}(x)-f_{\mathbb{L}}(x))^{2}d\nu (x). \label{L22}
\end{equation}

\bigskip \noindent (2) The family of Renyi's divergence measures indexed by $\alpha \neq 1$, $\alpha>0$, known under the name of Renyi-$\alpha$ : 
\begin{equation}
\mathcal{D}_{R,\alpha }(\mathbb{Q},\mathbb{L)=}\frac{1}{\alpha -1}\log
\left( \int_{\mathbb{R}^{d}}f_{\mathbb{Q}}^{\alpha }(x)f_{\mathbb{L}%
}^{1-\alpha }(x)d\nu (x)\right). \label{renyi}
\end{equation}

\bigskip \noindent (3) The family of Tsallis divergence measures indexed by $\alpha \neq 1$, $\alpha>0$, also known under the name of Tsallis-$\alpha$ : 
\begin{equation}
\mathcal{D}_{T,\alpha }(\mathbb{Q},\mathbb{L)}=\frac{1}{\alpha -1}\left(
\int_{\mathbb{R}^{d}}f_{\mathbb{Q}}^{\alpha }(x)f_{\mathbb{L}}^{1-\alpha
}(x)-1\right) d\nu (x);  \label{tsal}
\end{equation}

\bigskip \noindent (4) The Kullback-Leibler divergence measure
\begin{equation}
\mathcal{D}_{KL}(\mathbb{Q},\mathbb{L})=\int_{\mathbb{R}^{d}}f_{\mathbb{Q}%
}(x)\text{ }\log (f_{\mathbb{L}}(x)/f_{\mathbb{Q}}(x))\text{ }d\nu (x).
\label{kull1}
\end{equation}

\bigskip \noindent The latter, the Kullback-Leibler measure, may be interpreted as a limit case
of both the Renyi's family and the Tsallis' one by letting $\alpha
\rightarrow 1$. As well, for $\alpha $ near 1, the Tsallis family may be
seen as derived from $\mathcal{D}_{R,\alpha }(\mathbb{Q},\mathbb{L)}$ based on the first order expansion of the logarithm
function in the neighborhood of the unity. \newline

\noindent From this small sample of divergence measures, we may give the following remarks.\\

\noindent (a) The $L_{2}^{2}$-divergence measure is both an application and a metric on $\mathcal{P}^2$, where $\mathcal{P}$ is the class of probability measures on $\mathbb{R}^d$ such that
$$
\int_{\mathbb{R}^{d}} f^{2}_{\mathbb{Q}}(x) \ d\nu (x) <+\infty.
$$

\noindent (b) For both the Renyi and the Tsallis families, we may have integrability problems and lack of symmetry. For $d=1$,
it is clear from the very form of these divergence measures that we do not have symmetry, unless for the special case where $\alpha=1/2$. Next, consider two real random variables $X$ and $Y$ following gamma laws with respective shape parameters $(a,b)\in ]0,+\infty[^2$ and $(c,d)\in ]0,+\infty[^2$. Here naturally, we use \textit{pdf}'s with respect to the Lebesgue measure $\lambda $ on $\mathbb{R}$. Both families are build on the following functional 
\begin{equation*}
\mathcal{I}_{\alpha}(\mathbb{P}_{X},\mathbb{P}_{Y}, )=\int f_{\mathbb{P}
_{X}}^{\alpha }(x)f_{\mathbb{P}_{Y}}^{1-\alpha }(x)d\lambda (x)
\end{equation*}

\noindent which amounts, in this case, to 

\begin{equation*}
\mathcal{I}_{\alpha }(\mathbb{P}_{X},\mathbb{P}_{Y})=\frac{(\Gamma
(a+b)^{\alpha }}{(\Gamma (a))^{\alpha }(\Gamma (b))^{\alpha }}\frac{(\Gamma
(c+d))^{1-\alpha }}{(\Gamma (c))^{1-\alpha }(\Gamma (d))^{1-\alpha }}\frac{%
\Gamma (\alpha a+(1-\alpha )c)\Gamma (\alpha b+(1-\alpha )d)}{\Gamma (\alpha
a+(1-\alpha )c+\alpha b+(1-\alpha )d)}.
\end{equation*}

\bigskip \noindent This quantity is finite if and only if

\begin{equation*}
\alpha a+(1-\alpha )c\geq 0\text{ and }\alpha b+(1-\alpha )d\geq 0.
\end{equation*}

\bigskip \noindent From this sample tour, we have to be cautious, when speaking about divergence measures as applications and/or metrics. In the most general case, we have to consider the divergence measure between two specific probability measures as a number or a real parameter.\\

\bigskip \noindent Originally, divergence measures came as extensions and developments of information theory that was first set for discrete probability measures. In such a situation, the boundedness of these discrete probability measures above zero and below $+\infty$ was guaranteed. That is, the following assumption holds :\\

\noindent \textbf{Boundedness Assumption (BD)}. There exist two finite numbers $0<\kappa _{1}<\kappa _{2}<+\infty $ such that 
\begin{equation}
\kappa _{1}\leq f_{\mathbb{Q}},f_{\mathbb{L}}\leq \kappa _{2}.  \label{BD}
\end{equation}

\noindent If Assumption (\ref{BD}) holds, we do not have to worry about integrability problems, especially for Tsallis, Renyi and Kullback-Leibler measures, in the computations arising in the estimation theories. But, in the generalized context where are used arbitrary density functions with respect to some measure $\nu$, such an assumption is not that automatic. This explains why Assumption (\ref{BD}) is systematically used in a great number of works in that topic, for example, in \cite{singh}, \cite{kris}, \cite{hall}, to cite a few.\\

\noindent To ensure that Assumption (\ref{BD}) is fulfilled, it may be instrumental to restrict the computation of the integral used in the divergence measure to a \textbf{compact} domain $D$ such that 
$$
D_{1}=\int_{D}f_{\mathbb{Q}}(x)d\nu(x)>0 \text{ and } D_{2}=\int_{D}f_{\mathbb{P}}(x)d\nu(x)>0,
$$

\bigskip
\noindent and next, to appeal to the :\\

\noindent \textbf{Modified Boundedness Condition} : There exist  $0<\kappa _{1}<\kappa _{2}<+\infty$ and a \textbf{compact} domain $D$ as large as possible such that

\begin{equation}
\kappa _{1}\leq f_{\mathbb{Q}} 1_{D},f_{\mathbb{L}}1_{D}\leq \kappa _{2}.  \label{BDE}
\end{equation}

\bigskip \noindent This implies that the modified divergence measure, denoted by  $\mathcal{D}^{(m)}$, is applied to the modified \textbf{pdf}'s :

$$
f^{(m)}_{\mathbb{Q}}=D_{1}^{-1} f_{\mathbb{Q}} \text{ and } f^{(m)}_{\mathbb{P}}=D_{2}^{-1}f_{\mathbb{P}}.
$$

\bigskip \noindent Based of this technique, that we apply in case of integrability problems, we will suppose, when appropriate, \textbf{that Assumption (\ref{BD}) holds on a compact set $D$}.\\

\noindent Although we are focusing on the aforementioned divergence measures in this paper, it is worth mentioning that there exist quite a few number of them. Let us cite for example the ones named after  : Ali-Silvey or $f$-divergence \cite{topsoe}, Cauchy-Schwarz, Jeffrey divergence (see \cite{evren}), Chernoff (See \cite{evren}) , Jensen-Shannon (See \cite{evren}). According to \cite{cichocki}, there is more than a dozen of different divergence measures in the literature.\\

\noindent Before coming back to our divergence measures estimation of interest, we want to highlight some important applications of them. Indeed, divergence has proven to be useful in applications. Let us cite a few of them :\\

\noindent (a) They heavily intervene in Information Theory and recently in Machine Learning.\\

\noindent (b) They be used as similarity measures in image registration or multimedia classification (see \cite{moreno}).\\

\noindent (c) They are also used as loss functions in evaluating and optimizing the performance of density estimation
methods (see \cite{hall}).\\

\noindent (d) Divergence estimates can also be used to determine sample sizes required to achieve given performance levels in hypothesis testing.\\

\noindent (e) There has been a growing interest in applying divergence to various fields of science and engineering for the purpose of estimation,
classification, etc. (See \cite{bhattacharya}, \cite{liu1}).\\

\noindent (f) Divergence also plays a central role in the frame of large deviations results including the asymptotic rate of decrease of error probability in binary hypothesis testing problems.\newline

\noindent (g) The estimation of divergence between the samples drawn from unknown distributions gauges the distance between those distributions.
Divergence estimates can then be used in clustering and in particular for deciding whether the samples come from the same distribution by comparing
the estimate to a threshold.\\

\noindent (h) Divergence gauges how differently two random variables are distributed and it provides a useful measure of discrepancy between
distributions. In the frame of information theory , the key role of divergence is well known.\\

\noindent The reader may find more applications and descriptions in the following papers :  \cite{kullback},\cite{fukunaga}, 
\cite{cardoso}, \cite{ojala}, \cite{hastie}, \cite{buccigrossi}, \cite{moreno},\cite{macKay}.\\

\bigskip \noindent In the next subsection, we describe the frame in which we place the estimation problems we deal in this paper.\\

\subsection{Statistical Estimation} \label{subsec_intro_estim}$ $\\

\noindent The divergence measures may be applied to two statistical problems among others.\\

\noindent \textbf{(A)} First, it may be used as a fitting problem as described here. Let $X_{1},X_{2},....$ a sample from $X$ with an unknown probability distribution $\mathbb{P}_{X}$ and we want to test the hypothesis that $\mathbb{P}_{X}$ is equal to a known and fixed probability $\mathbb{P}_{0}.$ Theoretically, we can answer this question by estimating a divergence measure $\mathcal{D}(\mathbb{P}_{X},\mathbb{P}_{0})$ by a plug-in estimator $\mathcal{D}(\mathbb{P}_{X}^{(n)},\mathbb{P}_{0})$ where, for each $n\geq 1$, $\mathbb{P}_{X}$ is replaced by an estimator $\mathbb{P}_{X}^{(n)}$ of the probability law, which is based on sample $X_1$, $X_2$, ..., $X_n$, to be precised.\\

\noindent From there establishing an asymptotic theory of $\Delta _{n}=\mathcal{D}(%
\mathbb{P}_{X}^{(n)},\mathbb{P}_{0})-\mathcal{D}(\mathbb{P}_{X},\mathbb{P}%
_{0})$ is thought to be necessary to conclude.\\

\noindent \textbf{(B)} Next, it may be used as tool of comparing for two distributions. We may have two samples and wonder whether they come from the same probability measure. Here, we also
may two different cases.\\

\noindent \textbf{(B1)} In the first, we have two independent samples $%
X_{1},X_{2},....$ and $Y_{1},Y_{2},....$ respectively from a random variable 
$X$ and $Y.$ Here the estimated divergence $\mathcal{D}(\mathbb{P}_{X}^{(n)},%
\mathbb{P}_{Y}^{(m)})$, where $n$ and $m$ are the sizes of the available samples, is the natural estimator of $\mathcal{D}(\mathbb{P}%
_{X},\mathbb{P}_{Y})$ on which depends the statistical test of the hypothesis : $\mathbb{P}%
_{X}=\mathbb{P}_{Y}$.\\

\noindent \textbf{(B2)} But the data may also be paired $(X,Y)$, $(X_{1},Y_{2}),(X_{2},Y_{2}),...,$
that is $X_{i}$ and $Y_{i}$ are measurements of the same case $i=1,2,...$ In
such a situation, testing the equality of the margins $\mathbb{P}_{X}=\mathbb{P}%
_{Y} $ should be based on an estimator $\mathbb{P}_{(X,Y)}^{(n)}$ of the joint probability law of the couple $(X,Y)$ based of the
paired observations $(X_{i},Y_{i})$, $i=1,2,\ldots,n$.\\

\bigskip \noindent We did not encounter the approach (B2) in the literature. In the (B1) approach, almost all the papers used the same sample size, at the exception of \cite{poczos}, for the double-size estimation problem. In our view, the study case should rely on the available data so that using the same sample size may lead to a loss of information. To apply their method, one should take the minimum of the two sizes and then loose information. We suggest to come back to a general case and then study the asymptotic theory of $\mathcal{D}(\mathbb{P}_{X}^{(n)},\mathbb{P}_{Y}^{(m)})$ based on samples $X_{1},X_{2},..,X_n.$ and $Y_{1},Y_{2},...,Y_m$. In this paper, we will systematically use arbitrary samples sizes.\\

\noindent In the context of the situation (B1), there are several papers dealing with the estimation of the divergence measures. As we are concerned in this paper by the weak laws of the estimators, our review on that problematic did return only of a few results. Instead, the literature presented us many kinds of results on almost-sure efficiency of the estimation, with rates of convergences and laws of the iterated logarithm, $L^{p}$ ($p=1,2$) convergences, etc. To be precise, \cite{dakher} used recent techniques based on functional empirical process to provide a series of interesting rates of convergence of the estimators in the case of one-sided approach for the class de Renyi, Tsallis, Kullback-Leibler to cite a few. Unfortunately, the authors did not address the problem of integrability, taking text r=for granted that the divergence measures are finite. Although the results should be correct under the boundedness assumption \textit{BD} we described earlier, a new formulation in that frame would be welcome.\\

\noindent The paper of \cite{kris2} is exactly what we want to, except that is is concentrated of the $L^{2}$-divergence measure and used the Parzen approach. Instead, we will handle the most general case of $\phi$-divergence measure and will use the wavelets probability density estimators.\\ 

\bigskip \noindent In the context of the situation (B1), we may cite first the works of \cite{kris} and \cite{singh}. They both used divergence measures based on probability density functions and concentrated on Renyi-$\alpha $, Tsallis-$\alpha $ and Kullback-Leibler. \textbf{In the description of the results below, the estimated \textit{pfd}'s - f and g - are usually in a periodic H\H{o}lder class of a known smoothness $s$.}.

\bigskip \noindent Specifically, \cite{kris} defined Renyi and Tsallis estimators by correcting the plug-in estimator and established that, as long as 
$\mathcal{D}_{R,\alpha }(f,g) \geq c$ and $\mathcal{D}_{T,\alpha }(f,g) \geq c$, for some constant $c>0$, then
\begin{eqnarray*}
&& \mathbb{E}\left \vert \mathcal{D}_{R,\alpha} (f_n,g_n)-\mathcal{D}_{R,\alpha }(f,g)\right\vert  \leq  c \left( n^{-1/2}+n^{-\frac{3s}{2s+d}}
\right)\\
\text{and}&& \\
&& \mathbb{E}\left \vert \mathcal{D}_{T,\alpha }(f_n,g_n)-\mathcal{D}_{T,\alpha }(f,g)\right\vert  \leq  c \left( n^{-1/2}+n^{-\frac{3s}{2s+d}}
\right),
\end{eqnarray*}

\bigskip \noindent \cite{poczos} used a $k-$nearest-neighbor approach to prove that 
if $|\alpha -1|<k$, ($\alpha\neq 1)$ then 
\begin{eqnarray*}
&& \lim_{n,m\rightarrow \infty }\mathbb{E}\left[ \mathcal{D}_{T,\alpha }(f_n,g_m)-\mathcal{D}_{T,\alpha }(f,g)
\right] ^{2}=0\\
\text{and} &&\\
&&
\lim_{n,m\rightarrow \infty }\mathbb{E}\left( \mathcal{D}_{R,\alpha }(f_n,g_m)\right) =\mathcal{D}_{R,\alpha }(f,g).
\end{eqnarray*}

\bigskip \noindent There has been a recent interest in deriving convergence rates for divergence estimators (\cite{moon}, \cite{kris}). The rates are typically
derived in terms of smoothness $s$ of the densities : 

 \bigskip \noindent  The estimator of \cite{liu2} 
converges at rate $n^{-\frac{s}{s+d}}$, achieving the parametric rate when $s>d$.

\bigskip \noindent Similarly, \cite{sricharan} showed that when $s>d$ a $k$-nearest-neighbor style estimator achieves the rate $n^{-2/d}$ (in absolute error)
ignoring logarithmic factors. In a follow up work, the authors improved this result to $O(n^{-1/2})$ by using an
set of weak estimators, but they required $s > d$ orders of smoothness.

\bigskip \noindent \cite{singh} provided an estimator for R\'enyi$-\alpha $ divergences as well as general density functionals that uses
a \textit{mirror image} kernel density estimator. They obtained exponential inequalities for the deviation of the estimators from the true value.\\

 \bigskip \noindent \cite{kall} studied an $\varepsilon-$nearest neighbor estimator for the $L_2-$divergence that enjoys the same rate of
convergence as the projection-based estimator of \cite{kris}.\\

\bigskip \noindent The majority of the aforementioned articles worked with densities in H\H{o}lder classes, whereas our work applies for densities in the Besov classes.\\

\noindent Here, we will focus on divergence measures between absolutely continuous probability laws with respect to the Lebesgue measures. As well, our results applied to the approaches (A) and (B1) defined above. As a sequence, we estimate divergence measures by their plug-in counterparts, meaning that we replace the probability density functions (\textit{pdf}) in the expression of the divergence measure by a nonparametric estimators of the \textit{pdf}'s. From now, we have on our probability space, two independent sequences :\\

\noindent (-) a sequence of independent and identically distributed random variables with common \textit{pdf} $f_{\mathbb{P}_X}$ :
\begin{equation}
X_1, X_2, ...  \label{SX}
\end{equation}

\noindent (-) a sequence of independent and identically distributed random variables with common \textit{pdf} $g_{\mathbb{P}_Y}$ :
\begin{equation}
Y_1, Y_2, ...  \label{SY}
\end{equation}

\bigskip 
\noindent To make the notations more simple, we write

$$
f=f_{\mathbb{P}_X} \text{ and } g=f_{\mathbb{P}_Y}.
$$

\bigskip
\noindent We focus on using \textit{pdf}'s estimates provided by the wavelets approach. We will deal on the Parzen approach in a forthcoming study. So, we need to explain the frame in which we are going to express our results.\\

\bigskip \noindent We also wish to get, first, general laws for an arbitrary functional of the form 

\begin{equation} \label{defJ}
J(f,g)=\int_{D} \phi(f(x),g(x)) dx,
\end{equation}

\noindent where $\phi(x,y)$ is a measurable function of $(x,y) \in \mathbb{R}_{+}^{2}$ on which we will make the appropriate conditions. The results on the functional $J(f,g)$, which is also known under the name of $\phi$-divergence, will lead to those on the particular cases of the Renyi, Tsallis, and Kullback-Leibler measures.\\

\bigskip The rest of the paper is organized as follows. In the remainder part of this section, we describe the wavelets density estimators we will use alongside basic notation and assumptions. In Section 
\ref{sec2}, we will give our full results for the functional $J(f,g)$ both in one-sided and two-sided approaches.
In Section \ref{sec3}, we will particularize the results for specific measures we already described. The proofs are postponed in Section \ref{sec4}. Technical remarks are gathered in the Appendix Section
\ref{annexe}.

\subsection{Wavelets estimation of \textit{pdf}'s} \label{subsec_intro_wavelets}$ $\\

\noindent To begin with the wavelets theory and its statistical applications, we say that the wavelets setting involves two functions $\varphi $ and $\psi $ in $L_{2}(\mathbb{R)}$ respectively called \textit{farther} and \textit{mother} such that 
\begin{equation*}
\left\{ \varphi(. -k), \ 2^{j/2}\psi (2^{j}(.)-k),(j,k) \in \mathbb{Z}^{2}\right\},
\end{equation*}

\noindent is a orthonormal basis of $L_{2}(\mathbb{R)}$. We adopt the following notation, for $j\geq 0$, $k\in \mathbb{Z}$ :

$$
\varphi{j,k}=2^{j/2}\varphi (2^{j}(.)-k) \text{ and } \psi_{j,k}=2^{j/2}\psi (2^{j}(.)-k).
$$

\noindent Thus, any function $f$ in $L_{2}(\mathbb{R)}$ is characterized by its coordinates in the orthonormal basis, in the form
\begin{equation}
f= \sum_{k\in \mathbb{Z}} \alpha_{0,k}\varphi_{0,k}  + \sum_{k \in \mathbb{Z}} \sum_{j\geq 1}\beta_{j,k}\psi_{j,k} \label{decom_tot_wavelets}
\end{equation}

\noindent with for $j\geq 0$, $k\in \mathbb{Z}$,

$$
\alpha_{0,k}=\int_{\mathbb{R}} f(t)\varphi_{0,k}(t) \ dt \text{ and } \beta_{j,k}=\int_{\mathbb{R}} f(t)\psi_{j,k}(t) \ dt.
$$

\noindent For an easy introduction to the wavelets theory and to its applications to statistics, see for instance
\cite{hard}, \cite{daubechies}, \cite{bla}, etc. In this paper we only mention the unavoidable elements of this frame.\\

\noindent Based on the orthonormal basis defined below, the following Kernel function is introduced
$$
\mathbb{R}^{2}\ni (x,y) \mapsto K(x,y)=\sum_{k\in \mathbb{Z}}\varphi (x-k) \varphi(y-k).
$$ 

\noindent For any $j\geq 1$ fixed, called a resolution level, we define  
\begin{equation*}
K_{j}(x,y)=2^{j}K(2^{j}x,2^{j}y)
\end{equation*}

\noindent and for measurable function $h$,  we define the operator projection $K_{j}$ of $h$
onto the space $V_{j}$ of $L_2(\mathbb{R})$ (spanned by $2^{j/2}\varphi (2^{j}(.)-k)$), by 
\begin{equation*}
\mathbb{R} \ni x \mapsto K_{j}(h)(x)=\int K_{j}(x,y)h(y)dy. 
\end{equation*}

\noindent Therefore we can write, for all  $x\in \mathbb{R}$,

\begin{eqnarray}  \label{kj1}
K_{j}(h)(x)&=&2^{j}\int K(2^{j}x,2^{j}y)h(y)dy  \notag \\
&=& 2^{j}\int \sum_k \varphi(2^{j}x-k)\varphi(2^{j}y-k)h(y)dy. \label{KJ2}
\end{eqnarray}

\noindent In the frame of this wavelets theory, for each $n\geq 1$, we fix the resolution level depending on $n$ and denoted by $j=j_n$, and we use 
the following estimator of the \textit{pdf} $f$ associated to $X$, based on the sample of size $n$ from $X$, as defined in (\ref{SX}),
\begin{equation}  \label{fn}
f_{n}(x)=\frac{1}{n}\sum_{i=1}^{n}K_{j_{n}}(x,X_{i}).
\end{equation}

\noindent As well, in a two samples problem, we will estimate the \textit{pdf} $g$ associated to $Y$, based of a sample of size $n$ from $Y$, as defined in (\ref{SY}),  by
 
\begin{equation}  \label{gn}
g_{n}(x)=\frac{1}{n}\sum_{i=1}^{n}K_{j_{n}}(x,Y_{i}).
\end{equation}

\noindent The aforementioned estimator is known under the name linear wavelets estimators.\\

\noindent Before we give the main assumptions on the wavelets we are working, we have to define the concept of weak differentiation. Denote by $\mathcal{D}(\mathbb{R})$ the class of functions from $\mathbb{R}$ to $\mathbb{R}$ with compact support and infinitely differentiable. A function $f: \mathbb{R}  \rightarrow \mathbb{R}$ is weak differentiable if and only if there exists a function $g: \mathbb{R}  \rightarrow \mathbb{R}$ locally integrable (on compact sets) such that, for any 
$\phi \in \mathcal{D}(\mathbb{R})$, we have

$$
\int f(u) \phi^{\prime}(u) du= - \int g(u) \phi(u) du.
$$

\noindent In such a case, $g$ is called the weak derivative function of $f$ and denoted $f^{[1]}$. If the first weak derivative has itself a weak derivative, ans so forth up to the $p-1$-th derivative, we get the $p$-th derivative function $f^{[p]}$. Now we may expose the four assumptions we require on the wavelets.\\

\begin{assumption} \label{C1} . The wavelets $%
\varphi $ and $\psi $ are bounded and have compact support and either (i)
the father wavelet $\varphi $ has weak derivatives up to order $T$ in $L_{p}(%
\mathbb{R)}\,(1\leq p\leq \infty$ ) or (ii) the mother wavelet $\psi $
associated to $\varphi$ satisfies $\int x^{m}\psi (x)dx=0$ for all $%
m=0,\ldots ,T.$
\end{assumption}

\noindent and

\begin{assumption}  \label{C2} $\varphi : \mathbb{R}%
\rightarrow \mathbb{R}$ is of bounded $p$-variation for some $1\leq p<\infty 
$ and vanishes on $(B_{1},B_{2}]^{c\text{ }}$ for some $-\infty
<B_{1}<B_{2}<\infty .$
\end{assumption}

\noindent Wavelets generators with compact supports are available in the literature. We may cite those named after Daubechies, Coiflets
and Symmlet (See \cite{hard}). The cited generators fulfill our two main assumption.\\

\noindent Under \textbf{Assumption} \ref{C2}, the summation over $k$, in  (\ref{KJ2}), is
finite since only a number of the terms in the summation are non zeros (see \cite{gine01}).\newline

\noindent The third assumption concerns the resolution level we choose. We set for once an increasing sequence $(j_n)_{n\geq 1}$ such that

\begin{assumption} \label{C4}  There exists a non-negative symmetrical and continuous function $\Phi(t)$ of $t\in \mathbb{R}$ with a compact support $\mathcal{K}$ such that :
$$
\forall (x,y)\in \mathbb{R}^2, \left\vert K(x,y) \right\vert \leq \Phi(x-y).
$$
\end{assumption}

\begin{assumption}
\label{C3}  $\lim_{n\rightarrow +\infty} n^{-1/4}2^{j_{n}}=1$.
\end{assumption}

\noindent By the way, we have as $n \rightarrow \infty $, and

\begin{eqnarray}
\sqrt{\frac{j_{n}2^{j_{n}}}{n}}+2^{-tj_{n}} &\approx &\sqrt{\frac{1}{4\log 2}%
\frac{\log n}{n^{3/4}}}+n^{-t/4}\rightarrow 0, \ \ \forall t>0  \label{jn} \\
\frac{j_{n}}{\log \log n} &\rightarrow &\infty \text{\ \ \ and \ \ }\sup_{n\geq n_{0}}(j_{2n}-j_{n})=\frac{1}{4}\text{.}  \notag
\end{eqnarray}

\bigskip \noindent These conditions allow the use the \cite{gine01}'s results.\\

\noindent We also denote 
\begin{eqnarray}  \label{abcn}
&&a_{n}=\left\Vert f_{n}-f\right\Vert _{\infty },\text{ }b_{n}=\left\Vert g_{n}-g\right\Vert _{\infty}, \ n\geq 1 \label{abcn}\\
&&c_{n}=a_{n}\vee b_{n}, \text{ }c_{n,m}=a_{n}\vee b_{m}, n\geq 1,\ m\geq 1, \notag\\
&&c^{\ast}_{n,m}=c_{n,m} \vee c_{m,n}, \ n\geq 1,\ m\geq 1. \notag
\end{eqnarray}

\bigskip
\noindent where $\left\Vert h\right\Vert _{\infty }$stands for $\sup_{x\in
D(h)}\left\vert h(x)\right\vert$, and $D(h)$ is the domain of application of $h$.\\

\noindent In the sequel we suppose the densities $f$ and $g$ belong to the Besov space $\mathcal{B}_{\infty ,\infty }^{t}\left( \mathbb{R}\right)$. We will say a word of simple conditions under which our \textbf{pdf}'s do belong to such spaces.\\

\noindent Suppose that the densities $f$ and $g$ belong to $\mathcal{B}_{\infty ,\infty
}^{t}\left( \mathbb{R}\right) $, that $\varphi $ satisfies \textbf{Assumption} \ref{C2}, and $\varphi ,\psi $ satisfy \textbf{Assumption} \ref{C1}. Then Theorem 3  \cite{gine01} implies that the rates of convergence $a_{n},$ $b_{n}$ and $c_{n}$ are of the form
\begin{eqnarray*}
O\left( \sqrt{\frac{1}{4\log 2} \frac{\log n}{n^{3/4}}} +n^{-t/4}\right)
\end{eqnarray*}

\noindent almost-surely and converge all to zero at this rate (with $0<t<T$).\\

\bigskip \noindent In order to establish the asymptotic normality of the divergences estimators, we need this key tool concerning the wavelets empirical process denoted by $\mathbb{G}_{n,X}^{w}(h)$, where $h\in \mathcal{B}_{\infty ,\infty }^{t}\left( \mathbb{R}\right)$ and defined as follows by
 
\begin{eqnarray*}
\mathbb{G}_{n,X}^{w}(h)&=&\sqrt{n}\left( \mathbb{P}_{n,X}^{w}-\mathbb{E}%
_{X}\right)(h),
\end{eqnarray*}

\noindent where $\mathbb{P}_{n,X}^{w}(h)=\mathbb{P}_{n,X}\left( K_{j_{n}}(h)\right)= 
\frac{1}{n}\sum_{i=1}^{n}K_{j_{n}}(h)(X_{i})$ and $\mathbb{E}_X(h)=\int h(x)f(x)dx $ denotes the expectation of the measurable function $h$ with respect to the probability distribution function $\mathbb{P}_X$. The superscript $w$ refers to \textit{wavelets}. We have
 
\begin{equation}  \label{gnx1}
\mathbb{G}_{n,X}^{w}(h)=\sqrt{n}\int (f_{n}(x)-f(x))h(x)dx
\end{equation}

\noindent since, by Fubini's Theorem, 
\begin{eqnarray*}
\sqrt{n}\left( \mathbb{P}_{n,X}^{w}-\mathbb{E}_{X}\right)(h)&=&\sqrt{n}%
\left( \frac{1}{n}\sum_{i=1}^{n}K_{j_{n}}(h)(X_{i})-\int f(x)h(x)dx\right) \\
&=&\sqrt{n}\left(\frac{1}{n}\sum_{i=1}^{n}\int K_{j_{n}}(x,X_{i})h(x)dx-\int
f(x)h(x)dx \right) \\
&=&\sqrt{n}\int \left( \frac{1}{n}\sum_{i=1}^{n}K_{j_{n}}(x,X_{i})-f(x)%
\right) h(x)dx \\
&=& \sqrt{n}\int (f_{n}(x)-f(x))h(x)dx.
\end{eqnarray*}

\bigskip \noindent We are ready to give our results on the functional $J$ introduced in Formula (\ref{defJ}).

\section{RESULTS} \label{sec_mainResults} \label{sec2}

\subsection{Main Results} \label{subsec_mainResults_main}$ $\\

\noindent Here, we present a general asymptotic theory of a class of divergence measures estimators including the Renyi and Tsallis families and the Kullback-Leibler ones.\\

\noindent Actually, we gather them in the $\phi$-divergence measure form. We will obtain a general frame from which we will derive a number of corollaries. The assumption (\ref{BD}) will be used in the particular cases to ensure the finiteness of the divergence measure as mentioned in the beginning of the article. However, in the general results, the assumption (\ref{BD}) is part of the general conditions.\\

\noindent We begin to state a result as a general tool for establishing asymptotic normality and related to the wavelets empirical process, which we will use for establishing the asymptotic normality of divergence measures.\\

\begin{theorem} \label{thgn} Given the $(X_{n})_{n\geq 1}$, defined in (\ref{SX}) such that $f\in \mathcal{B}_{\infty,\infty}^{t}(\mathbb{R})$ and let $f_{n}$ defined as \eqref{fn} and $\mathbb{G}_{n,X}^{w}$ defined as in \eqref{gnx1}. Then, under \textbf{Assumption} (\ref{C1}-\ref{C4}) and for any bounded $h$, defined on $D$, belonging to $\mathcal{B}_{\infty ,\infty }^{t}\left( \mathbb{R}\right)$, we have 
\begin{equation*}
\sigma_{h,n}^{-1} \mathbb{G}_{n,X}^{w}(h)\rightsquigarrow \mathcal{N}(0,1)\ \text{
as }n\rightarrow \infty,
\end{equation*}

\bigskip
\noindent where we have
\begin{eqnarray}  \label{sig}
\sigma _{h,n}^{2}=\mathbb{E}_{X}\left(K_{j_{n}}(h)(X)%
\right)^{2}-\left(\mathbb{E}_{X}(K_{j_{n}}(h)(X)\right)^{2}  \rightarrow \mathbb{V}ar (h(X)) \ \text{
	as }\ n\rightarrow \infty. \notag
\end{eqnarray}

\end{theorem}

\noindent Based on that result which will be proved later, we are going to state all results of the functional $J$ defined in Formula \ref{defJ}, regarding its almost-sure and Gaussian asymptotic behavior. Let us begin by some notations. Let us assume that $\phi$ have continuous second order partial derivatives defined as follows : 
\begin{equation*}
\phi _{1}^{(1)}(s,t)=\frac{\partial \phi }{\partial s}(s,t),\text{ }\phi
_{2}^{(1)}(s,t)=\frac{\partial \phi }{\partial t}(s,t)
\end{equation*}

\noindent and

\begin{equation*}
\phi _{1}^{(2)}(s,t)=\frac{\partial ^{2}\phi }{\partial s^{2}}(s,t),\text{ }%
\phi _{2}^{(2)}(s,t)=\frac{\partial ^{2}\phi }{\partial t^{2}}(s,t),\text{ }%
\phi _{1,2}^{(2)}(s,t)=\phi _{2,1}^{(2)}(s,t)=\frac{\partial ^{2}\phi }{%
\partial s\partial t}(s,t).
\end{equation*}

\bigskip \noindent Define the functions $h_{i}$, $i=1,\ldots4$ : 
\begin{equation*}
h_{1}(x)=\phi _{1}^{(1)}(f(x),g(x)), \ h_{2}(x)=\phi _{2}^{(1)}(f(x),g(x)),
\end{equation*}

\begin{equation*}
h_{3}(x)=\phi _{1}^{(1)}(g(x),f(x))\text{ and }h_{4}(x)=\phi _{2}^{(1)}(g(x),f(x))
\end{equation*}

\noindent Set 
 
\begin{equation*}
A_{1}=\int_{D} \left\vert h_{1}(x)\right\vert dx\text{ }\text{\ \ and \ } A_{2}=\int_{D} \left\vert h_{2}(x)\right\vert dx
\end{equation*}

\noindent and

\begin{equation*}
A_{3}=\int_{D} \left\vert h_{3}(x)\right\vert dx\text{ }\text{\ \ and \ } A_{4}=\int_{D} \left\vert h_{4}(x)\right\vert dx.
\end{equation*}

\bigskip \noindent We require the following general conditions.\\

\noindent C-$A$. All the constants $A_i$ are finite.\\

\noindent C-$h$. All the functions $h_i$ used in the theorem below are bounded and lie in a Besov space $\mathcal{B}^{t}_{\infty \infty}$ for some $t$ such that $t>1/2$.\\

\noindent C1-$\phi$. The following integrals 

\begin{equation*}
\int \left\{ |\phi _{1}^{(1)}(f(x),g(x))|+|\phi
_{2}^{(1)}(f(x),g(x))|\right\} dx<+\infty .
\end{equation*}

\noindent are finite.\\

\bigskip \noindent C2-$\phi$. For any measurable sequences of functions $\delta _{n}^{(1)}(x),$
$\delta _{n}^{(2)}(x),$ $\rho _{n}^{(1)}(x),$ and $\rho _{n}^{(2)}(x)$ of $%
x\in D,$ uniformly converging to zero, that is
\begin{equation*}
\max_{i=1,2,\text{ }j=1,2}\sup \left\{ \left\vert \delta
_{n}^{(i)}(x)\right\vert +\left\vert \rho _{n}^{(j)}(x)\right\vert \right\}
<+\infty ,
\end{equation*}

\noindent we have as $n\rightarrow \infty$
\begin{equation}
\int_{D} \phi _{1}^{(2)}\left( f(x)+\delta _{n}^{(1)}(x),g(x)\right)
dx\rightarrow \int_{D} \phi _{1}^{(2)}(f(x),g(x))dx,  \label{CCS1}
\end{equation}

\begin{equation}
\int_{D} \phi _{2}^{(2)}\left( f(x),g(x)+\delta _{n}^{(2)}(x)\right)
dx\rightarrow \int_{D} \phi _{2}^{(2)}(f(x),g(x))dx,  \label{CCS2}
\end{equation}

\noindent and 
\begin{equation}
\int_{D} \phi _{1,2}^{(2)}\left( f(x)+\rho _{n}^{(1)}(x),g(x)+\rho
_{n}^{(2)}(x)\right) dx\rightarrow \int_{D} \phi _{1,2}^{(2)}(f(x),g(x))dx.  \label{CCD}
\end{equation}

\bigskip 
\begin{remark} $ $\\
\noindent (a) To check C-$h$, we may use the criteria we state in the Appendix Section \ref{annexe}, especially when dealing with usual distributions.\\

\noindent (b) The conditions in C2-$\phi$ may be justified by the Dominated Convergence Theorem or the monotone Convergence Theorem or from other limit theorems. We may either
express conditions on the general function $\phi$ under which these results hold true. But here, we choose to state the final results and next, to check them for particular cases, in which we may use convergence theorems.
\end{remark}

\bigskip \noindent  Based on \eqref{fn} and \eqref{gn}, we will use the following estimators 
\begin{eqnarray*}
J(f_{n},g)&=&\int_{D}\phi (f_n(x),g(x))dx, \text{\ \ \ \ }%
J(f,g_{n})=\int_{D}\phi (f(x),g_n(x))dx, \\
\text{\ \ and \ \ } J(f_{n},g_{n})&=&\int_{D}\phi (f_{n}(x),g_n(x))dx.
\end{eqnarray*}

\bigskip \noindent  Here are our main results.\\

\bigskip \noindent \textbf{I - Statements of the main results}.\\

\noindent The first concerns the almost sure efficiency of the estimators.\\

\begin{theorem} \label{thJ12} Under the assumptions \ref{C1}-\ref{C4}, C-$A$, C-$h$, C1-$\phi$, C2-$\phi$ and (BD),  we have  
\begin{eqnarray}
&&\limsup_{n\rightarrow +\infty } \frac{\left\vert
J(f_{n},g)-J(f,g)\right\vert}{a_{n}} \leq A_{1},\text{a.s}  \label{thJ12c1}
\\
&& \limsup_{n\rightarrow +\infty } \frac{\left\vert
J(f,g_{n})-J(f,g)\right\vert}{b_{n}} \leq A_{2},\text{a.s}  \label{thJ12c2}
\\
&& \limsup_{(n,m) \rightarrow (+\infty,+\infty)}\left\vert \frac{J(f_{n},g_{m})-J(f,g)}{%
c_{n,m}}\right\vert \leq A_{1}+A_{2} \text{\ \ a.s }  \label{thJ22c1}
\end{eqnarray}

\bigskip
\noindent where $a_{n}$, $b_n$ and $c_{n}$ are as in \eqref{abcn}.
\end{theorem}

\bigskip \bigskip
\noindent The second concerns the asymptotic normality of the estimators.\\

\begin{theorem} \label{thJ22} Under the assumptions \ref{C1}-\ref{C4}, C-$A$, C-$h$, C1-$\phi$, C2-$\phi$ and (BD),  we have
\begin{equation}
\sqrt{n}(J(f_{n},g)-J(f,g))\rightsquigarrow \mathcal{N}\left( 0,\mathbb{V}ar(h_1(X)) \right) ,\text{ as }n\rightarrow +\infty  \label{thJ12n1}
\end{equation}

\begin{equation}
\sqrt{n}(J(f,g_{n})-J(f,g))\rightsquigarrow \mathcal{N}\left( 0, \mathbb{V}ar(h_2(Y))\right) ,\text{ as }n\rightarrow +\infty  \label{thJ12n2}
\end{equation}

\noindent and  as $n\rightarrow +\infty$ and $m\rightarrow +\infty$, 

\begin{equation}
\left(\frac{nm}{m\mathbb{V}ar(h_{1}(X)+n\mathbb{V}ar(h_{2}(Y))}\right)^{1/2} \biggr( J(f_{n},g_{m})-J(f,g)\biggr) \rightsquigarrow \mathcal{N}\left( 0,1\right).  \label{thJ22n1}
\end{equation}

\end{theorem}

\bigskip \noindent \textbf{II - Direct extensions}.\\

\noindent Quite a few number of divergence measures are not symmetrical. Among these non-symmetrical measures are some of the most interesting ones. For  such measures, estimators of the form $J(f_n,g)$, $J(f,g_n)$ and $J(f_n,g_n)$ are not equal to $J(g,f_n)$, $J(g_n,f)$ and $J(g_n,f_n)$ respectively.\\

\noindent In one-sided tests, we have to decide whether the hypothesis $f=g$, for $g$ known and fixed, is true based on data from $f$. In such a case, we may use the statistics one of the statistics $J(f_n,g)$ and  $J(g,f_n)$ to perform the tests. We may have information that allows us to prefer one of them over the other. If not, it is better to use both of them, upon the finiteness of both $J(f,g)$ and $J(g,f)$, in a symmetrized form as

\begin{equation}
J_{(s)}(f,g)=\frac{J(f,g)+J(g,f)}{2}.
\end{equation}

\noindent The same situation applies when we face double-side tests, i.e., testing $f=g$ from data generated from $f$ and from $g$.\\

\noindent \textbf{Asymptotic a.e. efficiency}. 

\begin{theorem}\label{thJs22} Under the assumptions \ref{C1}-\ref{C4}, C-$A$, C-$h$, C1-$\phi$, C2-$\phi$ and (BD), we have
\begin{eqnarray}
\label{thJs12c1}&&\ \ \ \ \ \ \ \ \ \limsup_{n\rightarrow +\infty} \frac{\left\vert J_{(s)}(f_n,g)-J_{(s)}(f,g)\right \vert}{a_{n}} \leq \frac{1}{2}\left( A_{1}+A_{4}\right)\ \  \text{ a.e.},\\
\label{thJs21c1}&&\ \ \ \ \ \ \ \ \ \limsup_{n\rightarrow +\infty} \frac{\left\vert J_{(s)}(f,g_n)-J_{(s)}(f,g)\right \vert}{a_{n}} \leq \frac{1}{2}\left( A_{2}+A_{3} \right)\ \  \text{ a.e.},\\
\label{thJs22c1}&&\ \ \ \ \ \ \ \ \ \limsup_{n\rightarrow +\infty} \frac{\left \vert J_{(s)}(f_n,g_n)-J_{(s)}(f,g)\right \vert}{c_{n}}\leq \frac{1}{2}\left( A_{1}+A_{2}+A_{3}+A_{4}\right), \text{ a.e.}.
\end{eqnarray}
\end{theorem}

\bigskip
\noindent \textbf{Asymptotic Normality}. \\

\noindent Denote

$$
\sigma_{1,4}^{2}=\mathbb{V}ar(h_{1}+h_{4})(X)) \ and \ \sigma_{2,3}^{2}=\sigma_{2,3}^{2}=\mathbb{V}ar(h_{2}+h_{3})(Y)).
$$

\noindent We have
\begin{theorem} Under the assumptions \ref{C1}-\ref{C4}, C-$A$, C-$h$, C1-$\phi$, C2-$\phi$ and (BD), we obtain

\begin{equation}
\sqrt{\frac{n}{\mathbb{V}ar(h_{1}+h_{4})(X)}}  \biggr(J_{(s)}(f_n,g)-J_{(s)}(f,g)\biggr) \rightsquigarrow \mathcal{N}(0, 1),
\end{equation}

\begin{equation}
\sqrt{\frac{n}{\mathbb{V}ar(h_{2}+h_{3})(X)}}  \biggr(J_{(s)}(f,g_n)-J_{(s)}(f,g)\biggr) \rightsquigarrow \mathcal{N}(0, 1).
\end{equation}

\noindent and

\begin{equation}
\left(\frac{nm}{m\sigma_{1,4}^{2}+n\sigma_{2,3}^{2}}\right)^{1/2} \biggr(J_{(s)}(f_n,g_m)-J_{(s)}(f,g)\biggr) \rightsquigarrow \mathcal{N}(0, 1).
\end{equation}
\end{theorem}

\bigskip \noindent \textbf{Remark} The proof of these extensions will not be given here, since they are straight consequences of the main results. As well, such considerations will not be made again for particular measures for the same reason.

\bigskip \noindent We are going to give special forms of these mains results in a number of corollaries. To handle the Renyi and the Tsallis families, we get general results on the functional
$$
\mathcal{I}(\alpha,f,g)=\int_{D} f^{\alpha}(x)g^{1-\alpha}(x) dx, \alpha>0.
$$

\noindent which is used by these families. In turn the treatment of both of them are derived from the $\mathcal{I}$ functional using the delta method. For all these particular cases, we do not give their proofs since the derive from the general cases by straightforward computations.

\newpage

\section{Particular cases} \label{sec3}

\noindent \textbf{A - Renyi and Tsallis families}.\\

\noindent These two families are expressed through the functional

$$
\mathcal{I}(\alpha,f,g)=\int_{D} f^{\alpha}(x)g^{1-\alpha}(x) dx, \alpha>0.
$$

\noindent which of the form of the $\phi$-divergence measure with
$$
\phi(x,y)=x^{\alpha}y^{1-\alpha}, (x,y)\in \{ (f(s),g(t)), \ (s,t)D^{2} \}.
$$

\noindent So we begin by : \\

\noindent \textbf{A - (a) - The asymptotic behavior of the functional $\mathcal{I}(\alpha)$}.\\

\noindent With a compact domain $D$ and under the boundedness assumptions, and under the condition that neither $f$ nor $g$ vanishes on $D$, all the conditions $C$-A, $C$-h, $C1-\phi$ and $C2-\phi$ hold. Particularly, $C2-\phi$ derives by the application of the Lebesgue Dominated Theorem. Besides $\phi$ has continuous partial derivatives bounded against zero, of all order. This entails that the functions $h_i$ are all in the required Besov spaces. Then under the conditions on the wavelets, we have the following results.

\noindent First, we have

\begin{corollary}Let Assumptions \ref{C1}-\ref{C4} hold, and let (BDE) be satisfied. Then for any $\alpha>0$, we have
$$
\limsup_{n\rightarrow +\infty} \frac{|\mathcal{I}(\alpha,f_n,g)-\mathcal{I}(\alpha,f,g)|}{a_n}\leq \alpha\int_{D} (f(x)/g(x))^{\alpha-1} dx=:A_1(\alpha),
$$

$$
\limsup_{n\rightarrow +\infty} \frac{|\mathcal{I}(\alpha,f,g_n)-\mathcal{I}(\alpha,f,g)|}{b_n}\leq |\alpha-1| \int_{D} (f(x)/g(x))^{\alpha} dx=:A_2(\alpha)
$$

\noindent and

$$
\limsup_{n\rightarrow +\infty, \ m\rightarrow +\infty} \frac{|\mathcal{I}(\alpha,f_n,g_m)-\mathcal{I}(\alpha,f,g)|}{c_n}\leq A_1(\alpha)+A_2(\alpha). 
$$
\end{corollary}

\noindent Denote

$$
\sigma_{1}^{2}(\alpha,f,g)=\alpha^2 \left( \left( \int_{D} g(x)(f(x)/g(x))^{2\alpha-1} dx\right) - \left( \int_{D} g(x)(f(x)/g(x))^{\alpha} dx\right)^2\right)
$$ 

\noindent and

$$
\sigma_{2}^{2}(\alpha,f,g)=(\alpha-1)^2 \left( \left( \int_{D} g(x)(f(x)/g(x))^{2\alpha} dx\right) - \left( \int_{D} g(x)(f(x)/g(x))^{\alpha} dx\right)^2\right)
$$ 

\noindent We have

\begin{corollary} Let Assumptions \ref{C1}-\ref{C4} hold, and let (BDE) be satisfied. Then for any $\alpha>0$, we have as $n\rightarrow +\infty$ and $m\rightarrow +\infty$,
$$
\sqrt{n} (\mathcal{I}(\alpha,f_n,g)-\mathcal{I}(\alpha,f,g)) \rightsquigarrow \mathcal{N}(0, \sigma_{1}^{2}(f,g)),
$$

$$
\sqrt{n} (\mathcal{I}(\alpha,f,g_n)-\mathcal{I}(\alpha,f,g)) \rightsquigarrow \mathcal{N}(0, \sigma_{2}^{2}(f,g)),
$$

\noindent and

$$
\left(\frac{mn}{n\sigma_{2}^{2}(f,g)+m\sigma_{1}^{2}(f,g)}\right)^{1/2} \biggr(\mathcal{I}(\alpha,f_n,g_m)-\mathcal{I}(\alpha,f,g)\biggr) \rightsquigarrow \mathcal{N}(0, 1).
$$
\end{corollary}

\bigskip \noindent As to the symmetrized form

$$
\mathcal{I}_{s}(\alpha,f,g)=\frac{\mathcal{I}_{s}(\alpha,f,g)+\mathcal{I}_{s}(\alpha,g,f)}{2},
$$

\noindent we need the supplementary notations:

$$
A_3(\alpha,f,g)=\alpha \int_{D} (g(x)/f(x))^{\alpha-1} dx, \ , \ A_4(\alpha,f,g)=|\alpha-1| \int_{D} (g(x)/f(x))^{\alpha} dx,
$$

$$
\sigma_{1}^{2}(\alpha,f,g)=\alpha^2 \left( \left( \int_{D} g(x)(f(x)/g(x))^{2\alpha-1} dx\right) - \left( \int_{D} g(x)(f(x)/g(x))^{\alpha} dx\right)^2\right)
$$ 

\noindent 

$$
\ell_{1}(\alpha, x,y)=(y/x)^{\alpha}((1-\alpha)+\alpha (x/y)^{2\alpha-1}, \  \ell_{2}(\alpha, x,y)=(x/y)^{\alpha}((1-\alpha)+\alpha (y/x)^{2\alpha-1}, \ 
$$

$$
\sigma_{3}^{2}(\alpha,f,g)=\left( \int_D f(x)\ell_{1}(\alpha,f(x),g(x))^2 dx\right) -\left(\int_{D} f(x) \ell_{1}(\alpha,f(x),g(x)) dx\right)^2,
$$

\noindent

$$
\sigma_{4}^{2}(\alpha,f,g)=\left(\int_D g(x)\ell_{2}(\alpha,f(x),g(x))^2 dx \right)-\left(\int_{D} g(x)\ell_{2}(\alpha,f(x),g(x)) dx\right)^2.
$$

\noindent We have

\begin{corollary} Let Assumptions \ref{C1}-\ref{C4} hold and let (BDE) be satisfied. Then for any $\alpha>0$, 
$$
\limsup_{n\rightarrow +\infty} \frac{|\mathcal{I}_{(s)}(\alpha,f_n,g)-\mathcal{I}_{(s)}(\alpha,f,g)|}{a_n}\leq (A_1(\alpha)+A_4(\alpha))/2=:A_{1}^{(s)}(\alpha),
$$

$$
\limsup_{n\rightarrow +\infty} \frac{|\mathcal{I}_{(s)}(\alpha,f,g_n)-\mathcal{I}_{(s)}(\alpha,f,g)|}{b_n}\leq (A_2(\alpha)+A_3(\alpha))/2=:A_{2}^{(s)}(\alpha)
$$

\noindent and

$$
\limsup_{n\rightarrow +\infty, \ m\rightarrow +\infty} \frac{|\mathcal{I}_{(s)}(\alpha,f_n,g_m)-\mathcal{I}_{(s)}(\alpha,f,g)|}{c_{n,m}}\leq A_{1}^{(s)}(\alpha)+A_{2}^{(s)}(\alpha). 
$$
\end{corollary}

\noindent We also have
\begin{corollary} Let Assumptions \ref{C1}-\ref{C4} hold, and let (BDE) be satisfied. Then for any $\alpha>0$,  we have as $n\rightarrow +\infty$ and $m\rightarrow +\infty$,
$$
\sqrt{n} (\mathcal{I}(\alpha,f_n,g)-\mathcal{I}(\alpha,f,g)) \rightsquigarrow \mathcal{N}(0, \sigma_{3}^{2}(f,g)),
$$

$$
\sqrt{n} (\mathcal{I}(\alpha,f,g_n)-\mathcal{I}(\alpha,f,g)) \rightsquigarrow \mathcal{N}(0, \sigma_{3}^{2}(f,g)),
$$

\noindent and

$$
\left(\frac{mn}{n\sigma_{2}^{4}(f,g)+m\sigma_{3}^{2}(f,g)}\right)^{1/2} \biggr(\mathcal{I}_{(s)}(\alpha,f_n,g_m)-\mathcal{I}_{(s)}(\alpha,f,g)\biggr) \rightsquigarrow \mathcal{N}(0, 1).
$$
\end{corollary}

\bigskip \noindent
\noindent \textbf{A - (b) - Tsallis' Family}.\\

\noindent The treatment of the asymptotic behaviour of of the Renyi-$\alpha$, $\alpha>0$, $\alpha\neq 1$, is obtained from Part (A) by expansions and by the application of the delta method. We first remark that
$$
\mathcal{D}_{T,\alpha}(f,g)=\frac{\mathcal{I}(\alpha,f,g)}{\alpha-1}.
$$

\noindent We have the following results

\begin{corollary}Let Assumptions \ref{C1}-\ref{C4} hold, and let (BDE) be satisfied. Then for any $\alpha>0$, $\alpha\neq 1$, we have
$$
\limsup_{n\rightarrow +\infty} \frac{|\mathcal{D}_{T,\alpha}(f_n,g)-\mathcal{D}_{T,\alpha}(f,g)|}{a_n}\leq \frac{A_1(\alpha)}{|\alpha-1|}=:A_{T,\alpha,1},
$$

$$
\limsup_{n\rightarrow +\infty} \frac{|\mathcal{D}_{T,\alpha}(f,g_n)-\mathcal{D}_{T,\alpha}(f,g)|}{b_n}\leq \frac{A_2(\alpha)}{|\alpha-1|}=:A_{T,\alpha,2},
$$

\noindent and

$$
\limsup_{n\rightarrow +\infty, \ m\rightarrow +\infty} \frac{|\mathcal{D}_{T,\alpha}(f_n,g_m)-\mathcal{D}_{T,\alpha}(f,g)|}{a_n}\leq A_{T,\alpha,1}+A_{T,\alpha,2}.
$$
\end{corollary}

\noindent Denote

$$
\sigma_{T,1}^{2}(\alpha,f,g)=\frac{\sigma_{1}^{2}(\alpha,f,g)}{(\alpha-1)^2},\ and \ \sigma_{T,2}^{2}(\alpha,f,g)=\frac{\sigma_{2}^{2}(\alpha,f,g)}{(\alpha-1)^2}
$$

\noindent We have

\begin{corollary} Let Assumptions \ref{C1}-\ref{C4} hold, and let (BDE) be satisfied. Then for any $\alpha>0$, $\alpha\neq 1$ we have as $n\rightarrow +\infty$ and $m\rightarrow +\infty$,
$$
\sqrt{n} (\mathcal{D}_{R,\alpha}^{s}(f_n,g)-\mathcal{D}_{R,\alpha}(f,g)) \rightsquigarrow  \mathcal{N}(0, \sigma_{R,1}^{2}(\alpha,f,g)),
$$

$$
\sqrt{n} (\mathcal{D}_{T,\alpha}(f,g_n)-\mathcal{D}_{T,\alpha}(f,g)) \rightsquigarrow \mathcal{N}(0, \sigma_{T,2}^{2}(\alpha,f,g)),
$$

\noindent and

$$
\left(\frac{mn}{n\sigma_{T,2}^{2}(\alpha, f,g)+m\sigma_{T,1}^{2}(\alpha, f,g)}\right)^{1/2} \biggr(\mathcal{D}_{T,\alpha}(f_n,g_m)-\mathcal{D}_{T,\alpha}(f,g)\biggr) \rightsquigarrow \mathcal{N}(0, 1).
$$
\end{corollary}

\bigskip \noindent As to the symmetrized form

$$
\mathcal{D}^{(s)}_{R,\alpha}(f,g)(\alpha,f,g)=\frac{\mathcal{D}_{R,\alpha}(f,g)+\mathcal{D}_{R,\alpha}(g,f)}{2},
$$

\noindent we simply adapt the parameters obtained for the $A - (a)$. We have

$$
A_{T,\alpha,3}=A_{2,\alpha,3}/(|\alpha-1|), \ \ A_{T,\alpha,4}=A_{2,\alpha,4}/(|\alpha-1|).
$$

\noindent and

$$
\sigma_{T,3}^{2}(\alpha,f,g)=\sigma_{4}^{2}(\alpha,f,g)/(\alpha-1)^2, \ \ \sigma_{T,4}^{2}(\alpha,f,g)=\sigma_{4}^{2}(\alpha,f,g)/(\alpha-1)^2
$$

\noindent We have

\begin{corollary} Let Assumptions \ref{C1}-\ref{C4} hold, and let (BDE) be satisfied. Then for any $\alpha>0$, $\alpha\neq 1$,
$$
\limsup_{n\rightarrow +\infty} \frac{|\mathcal{D}_{T,\alpha}(f_n,g)^{(s)}-\mathcal{D}_{T,\alpha}(f,g)^{(s)}|}{a_n}\leq (A_{T,\alpha,1}+A_{T,\alpha,1})/2=:A_{T,\alpha,1}^{(s)}(\alpha),
$$

$$
\limsup_{n\rightarrow +\infty} \frac{|\mathcal{D}_{T,\alpha}(f,g_)^{(s)}-\mathcal{D}_{T,\alpha}(f,g_n)^{(s)}|}{b_n}\leq (A_{T,\alpha,2}+A_{T,\alpha,3})/2=:A_{T,\alpha,2}^{(s)}(\alpha)
$$

\noindent and

$$
\limsup_{n\rightarrow +\infty, \ m\rightarrow +\infty} \frac{|\mathcal{D}_{T,\alpha}(f_n,g_m)^{(s)}-\mathcal{D}_{T,\alpha}(f,g)^{(s)}|}{c_{n,m}}\leq A_{T,\alpha,1}^{(s)}+A_{T,\alpha,2}^{(s)}. 
$$
\end{corollary}

\noindent We also have
\begin{corollary} Let Assumptions \ref{C1}-\ref{C4} hold, and let (BDE) be satisfied. Then for any $\alpha>0$, $\alpha\neq 1$, we have as $n\rightarrow +\infty$ and $m\rightarrow +\infty$,
$$
\sqrt{n} (\mathcal{D}_{T,\alpha}(f_n,g)^{(s)}-\mathcal{D}_{T,\alpha}(f_n,g)^{(s)}) \rightsquigarrow \mathcal{N}(0, \sigma_{T,3}^{2}(f,g)),
$$

$$
\sqrt{n} (\mathcal{D}_{T,\alpha}^{s}(f_n,g)-\mathcal{D}_{T,\alpha}^{s}(f,g)) \rightsquigarrow \mathcal{N}(0, \sigma_{T,4}^{2}(f,g)),
$$

\noindent and

$$
\left(\frac{mn}{n\sigma_{2}^{T,4}(f,g)+m\sigma_{T,3}^{2}(f,g)}\right)^{1/2} \biggr(\mathcal{D}_{T,\alpha}^{s}(f_n,g_m)-\mathcal{D}_{T,\alpha}^{s}(f,g)\biggr) \rightsquigarrow \mathcal{N}(0, 1).
$$
\end{corollary}

\bigskip \noindent 
\noindent \textbf{A - (c) - Renyi's Family}.\\

\noindent The treatment of the asymptotic behaviour of of the Renyi-$\alpha$, $\alpha>0$, $\alpha\neq 1$, is obtained from Part (A) by expansions and by the application of the delta method. We first remark that
$$
\mathcal{D}_{R,\alpha}(f,g)=\frac{1}{\alpha-1} \log\left( \int_{D} f^{\alpha}(x)g^{1-\alpha}(x) dx\right)=\frac{\log(\mathcal{I}(\alpha,f,g))}{\alpha-1}.
$$

\noindent We have the following results

\begin{corollary}Let Assumptions \ref{C1}-\ref{C4} hold, and let (BDE) be satisfied. Then for any $\alpha>0$, $\alpha\neq 0$, we have
$$
\limsup_{n\rightarrow +\infty} \frac{|\mathcal{D}_{R,\alpha}(f_n,g)-\mathcal{D}_{R,\alpha}(f,g)}{a_n}\leq \frac{A_1(\alpha)}{|\alpha-1|\mathcal{I}(\alpha,f,g)}=:A_{R,\alpha,1},
$$

$$
\limsup_{n\rightarrow +\infty} \frac{|\mathcal{D}_{R,\alpha}(f,g_n)-\mathcal{D}_{R,\alpha}(f,g)}{b_n}\leq \frac{A_2(\alpha)}{|\alpha-1|\mathcal{I}(\alpha,f,g)}=:A_{R,\alpha,2},
$$

\noindent and

$$
\limsup_{n\rightarrow +\infty} \frac{|-\mathcal{D}_{R,\alpha}(f,g)}{a_n}\leq A_{R,\alpha,1}+A_{R,\alpha,2}.
$$
\end{corollary}

\noindent Denote

$$
\sigma_{R,1}^{2}(\alpha,f,g)=\frac{\sigma_{1}^{2}(\alpha,f,g)}{(\alpha-1)^2 \mathcal{I}(\alpha,f,g)^2},\ and 
\ \sigma_{R,2}^{2}(\alpha,f,g)=\frac{\sigma_{2}^{2}(\alpha,f,g)}{(\alpha-1)^2 \mathcal{I}\alpha,g,f)^2}
$$

\noindent We have

\begin{corollary} Let Assumptions \ref{C1}-\ref{C4} hold, and let (BDE) be satisfied. Then for any $\alpha>0$, $\alpha\neq 1$, we have as $n\rightarrow +\infty$ and $m\rightarrow +\infty$,
$$
\sqrt{n} (\mathcal{D}_{R,\alpha}^{s}(f_n,g)-\mathcal{D}_{R,\alpha}(f,g)) \rightsquigarrow \mathcal{N}(0, \sigma_{R,1}^{2}(\alpha,f,g)),
$$

$$
\sqrt{n} (\mathcal{D}_{R,\alpha}(f,g_n)-\mathcal{D}_{R,\alpha}(f,g)) \rightsquigarrow \mathcal{N}(0, \sigma_{R,2}^{2}(\alpha,f,g)),
$$

\noindent and

$$
\left(\frac{mn}{n\sigma_{R,2}^{2}(\alpha, f,g)+m\sigma_{R,1}^{2}(\alpha, f,g)}\right)^{1/2} \biggr(\mathcal{D}_{R,\alpha}(f_n,g_m)-\mathcal{D}_{R,\alpha}(f,g)\biggr) \rightsquigarrow \mathcal{N}(0, 1).
$$
\end{corollary}

\bigskip \noindent As to the symmetrized form

$$
\mathcal{D}^{(s)}_{R,\alpha}(f,g)(\alpha,f,g)=\frac{\mathcal{D}_{R,\alpha}(f,g)+\mathcal{D}_{R,\alpha}(g,f)}{2},
$$

\noindent we need the supplementary notations:

$$
A_{R,1}^{(s)}(\alpha,f,g)=\frac{1}{2|\alpha-1|}\left(\frac{A_{R,1}(f,g)}{2|\alpha-1|\mathcal{I}(f,g)}+\frac{A_{R,4}(f,g)}{2\mathcal{I}(g,f)}\right),
$$

$$
A_{R,2}^{(s)}(\alpha,f,g)=\frac{1}{2|\alpha-1|}\left(\frac{A_{R,2}(f,g)}{2|\alpha-1|\mathcal{I}(f,g)}+\frac{A_{R,3}(f,g)}{2\mathcal{I}(g,f)}\right),
$$

$$
\sigma_{1}^{2}(\alpha,f,g)=\alpha^2 \left( \left( \int_{D} g(x)(f(x)/g(x))^{2\alpha-1} dx\right) - \left( \int_{D} g(x)(f(x)/g(x))^{\alpha} dx\right)^2\right)
$$ 

\noindent 

$$
\ell_{R,1}(\alpha, x,y)=\frac{1}{2(\alpha-1)}\left(\frac{\alpha (x/y)^{\alpha-1}}{\mathcal{I}(f,g)}+\frac{((1-\alpha) (y/x)^{\alpha}}{\mathcal{I}(g,f)} \right),
$$

$$
\ell_{R,2}(\alpha, x,y)=\frac{1}{2(\alpha-1)}\left(\frac{\alpha (y/x)^{\alpha-1}}{\mathcal{I}(g,f)+\frac{(1-\alpha) (x/y)^{\alpha}}{\mathcal{I}(f,g)}} \right),
$$

$$
\sigma_{R,3}^{2}(\alpha,f,g)=\left( \int_D f(x)\ell_{R,1}(\alpha,f(x),g(x))^2 dx\right) -\left(\int_{D} f(x) \ell_{R,1}(\alpha,f(x),g(x)) dx\right)^2,
$$

\noindent

$$
\sigma_{R,4}^{2}(\alpha,f,g)=\left(\int_D g(x)\ell_{R,2}(\alpha,f(x),g(x))^2 dx \right)-\left(\int_{D} g(x)\ell_{R,2}(\alpha,f(x),g(x)) dx\right)^2.
$$

\noindent We have

\begin{corollary} Let Assumptions \ref{C1}-\ref{C4} hold, and let (BDE) be satisfied. Then for any $\alpha>0$, $\alpha\neq 1$,
$$
\limsup_{n\rightarrow +\infty} \frac{|\mathcal{D}_{R,\alpha}(f_n,g)^{(s)}-\mathcal{D}_{R,\alpha}(f_n,g)^{(s)}|}{a_n}\leq (A_{R,\alpha,1}+A_{R,\alpha,1})/2=:A_{R,\alpha, 1}^{(s)},
$$

$$
\limsup_{n\rightarrow +\infty} \frac{|\mathcal{D}_{R,\alpha}(f_n,g_n)^{(s)}-\mathcal{D}_{R,\alpha}(f,g)^{(s)}|}{b_n}\leq (A_{R,\alpha,2}+A_{R,\alpha,3})/2=:A_{R,\alpha, 2}^{(s)}
$$

\noindent and

$$
\limsup_{n\rightarrow +\infty, \ m\rightarrow +\infty} \frac{|\mathcal{D}_{R,\alpha}(f_n,g_m)^{(s)}-\mathcal{D}_{R,\alpha}(f_n,g)^{(s)}|}{c_{n,m}}\leq A_{R,\alpha, 1}^{(s)}+A_{R,\alpha, 2}^{(s)}. 
$$
\end{corollary}

\noindent We also have
\begin{corollary} Let Assumptions \ref{C1}-\ref{C2} hold, and let (BDE) be satisfied. Then for any $\alpha>0$, $\alpha\neq 1$, we have as $n\rightarrow +\infty$ and $m\rightarrow +\infty$,
$$
\sqrt{n} (\mathcal{D}_{R,\alpha}^{s}(f_n,g)-\mathcal{D}_{R,\alpha}^{s}(f,g)) \rightsquigarrow \mathcal{N}(0, \sigma_{R,3}^{2}(f,g)),
$$

$$
\sqrt{n} (\mathcal{D}_{R,\alpha}^{s}(f_n,g)-\mathcal{D}_{R,\alpha}^{s}(f,g)) \rightsquigarrow \mathcal{N}(0, \sigma_{R,4}^{2}(f,g)),
$$

\noindent and

$$
\left(\frac{mn}{n\sigma_{2}^{R,4}(f,g)+m\sigma_{R,3}^{2}(f,g)}\right)^{1/2} \biggr(\mathcal{D}_{R,\alpha}^{s}(f_n,g_m)-\mathcal{D}_{R,\alpha}^{s}(f,g)\biggr) \rightsquigarrow \mathcal{N}(0, 1).
$$
\end{corollary}

\bigskip \noindent \textbf{B- Kullback-Leibler Measure}.\\

\noindent Here we have

$$
\phi(x,y)=x \log(x/y), \ (x,y)\in \{ ((f(s),g(t)), \ (s,t)D^{2} \}.
$$

\noindent and the Kulback-Leibler Measure is defined by $\mathcal{D}_{KL}(f,g)=\int_{D} f(x) \log(f(x)/g(x) dx$.\\

\noindent The preliminary text of Part (A) is still valid. So, we have first :

\begin{corollary}Let Assumptions \ref{C1}-\ref{C4} hold, and let (BDE) be satisfied. Then  we have
$$
\limsup_{n\rightarrow +\infty} \frac{|\mathcal{D}_{KL}(f_n,g)-\mathcal{D}_{KL}(f,g)|}{a_n}\leq \int_{D} |1+\log(f(x)/g(x)| dx=:A_{DL,1}(f,g),
$$

$$
\limsup_{n\rightarrow +\infty} \frac{|\mathcal{D}_{KL}(f,g_n)-\mathcal{D}_{KL}(f,g)|}{b_n}\leq \int_{D} f(x)/g(x) dx=:A_{DL,2}(f,g),
$$

\noindent and

$$
\limsup_{n\rightarrow +\infty, m\rightarrow \infty} \frac{|\mathcal{D}_{KL}(f_n,g_m)-\mathcal{D}_{KL}(f,g)|}{c_n}\leq A_{DL,1}+A_{DL,2}.
$$
\end{corollary}

\noindent Denote

$$
\sigma_{DL,1}^{2}(f,g)= \left( \left( \int_{D} f(x)(1+\log(f(x)/g(x)^2 dx\right) - \left( \int_{D}  f(x)(1+\log(f(x)/g(x)dx\right)^2\right)
$$ 

\noindent and

$$
\sigma_{DL,2}^{2}-f,g)= \left( \left( \int_{D} f^{2}(x)/g(x) dx\right) - 1 \right)
$$ 

\bigskip \noindent We have

\begin{corollary} Let Assumptions \ref{C1}-\ref{C4} hold, and let (BDE) be satisfied. Then we have as $n\rightarrow +\infty$ and $m\rightarrow +\infty$,
$$
\sqrt{n} (\mathcal{D}_{KL}(f_n,g)-\mathcal{D}_{KL}(f,g)) \rightsquigarrow \mathcal{N}(0, \sigma_{DL,1}^{2}(f,g)),
$$

$$
\sqrt{n} (\mathcal{D}_{KL}(f,g_n)-\mathcal{D}_{KL}(f,g)) \rightsquigarrow \mathcal{N}(0, \sigma_{2}^{2}(f,g)),
$$

\noindent and

$$
\left(\frac{mn}{n\sigma_{DL,2}^{2}(f,g)+m\sigma_{DL,1}^{2}(f,g)}\right)^{1/2} \biggr(\mathcal{D}_{KL}(f_n,g_m)-\mathcal{D}_{KL}(f,g)) \rightsquigarrow \mathcal{N}(0, 1).
$$
\end{corollary}

\bigskip \noindent As to the symmetrized form

$$
\mathcal{I}_{s}(\alpha,f,g)=\frac{\mathcal{I}_{s}(\alpha,f,g)+\mathcal{I}_{s}(\alpha,g,f)}{2},
$$

\noindent we need the supplementary notations:

$$
A_{DL,3}(f,g)=\int_{D} |1+\log(g(x)/f(x)| dx, \ , \ A_{DL,4}(f,g)= \int_{D} g(x)/f(x) dx,
$$

$$
\ell_{DL,1}(x,y)=1-(y/x)+\log(x/y), \  \ell_{DL,2}(x,y)=1-(x/y)+\log(y/x), \ 
$$

$$
\sigma_{DL,3}^{2}(\alpha,f,g)=\left( \int_D f(x) \ell_{DL,1}(f(x),g(x))^2 dx\right) -\left(\int_{D} f(x) \ell_{DL,1}(f(x),g(x)) dx\right)^2,
$$

\noindent and

$$
\sigma_{DL,4}^{2}(f,g)=\left(\int_D g(x)\ell_{2}(\alpha,f(x),g(x))^2 dx \right)-\left(\int_{D} g(x)\ell_{2}(f(x),g(x)) dx\right)^2.
$$

\noindent We have

\begin{corollary} Let Assumptions \ref{C1}-\ref{C4} hold, and let (BDE) be satisfied. Then,
$$
\limsup_{n\rightarrow +\infty} \frac{|\mathcal{D}^{s}_{KL}(f_n,g)-\mathcal{D}^{s}_{KL}(f,g)|}{a_n}\leq (A_{DL,1}(f,g)+A_{DL,4})/2=:A_{DL,1}^{(s)}(f,g),
$$

$$
\limsup_{n\rightarrow +\infty} \frac{|\mathcal{D}^{s}_{KL}(f_n,g)-\mathcal{D}^{s}_{KL}(f,g)|}{b_n}\leq (A_{DL,2}(f,g)+A_{DL,3})/2=:A_{DL,2}^{(s)}(f,g)
$$

\noindent and

$$
\limsup_{n\rightarrow +\infty} \frac{|\mathcal{D}^{s}_{KL}(f_n,g)-\mathcal{D}^{s}_{KL}(f,g)|}{c_n}\leq A_{1}^{(s)}(A_{DL,1}^{(s)}(f,g)+A_{DL,2}^{(s)}(f,g)). 
$$
\end{corollary}

\noindent We also have
\begin{corollary} Let Assumptions \ref{C1}-\ref{C4} hold, and let (BDE) be satisfied. Then, we have as $n\rightarrow +\infty$ and $m\rightarrow +\infty$,
$$
\sqrt{n} (\mathcal{D}^{s}_{KL}(f_n,g)-\mathcal{D}^{s}_{KL}(f,g)) \rightsquigarrow \mathcal{N}(0, \sigma_{DL,3}^{2}(f,g)),
$$

$$
\sqrt{n} (\mathcal{D}^{s}_{KL}(f,g_n)-\mathcal{D}^{s}_{KL}(f,g)) \rightsquigarrow \mathcal{N}(0, \sigma_{DL,4}^{2}(f,g)),
$$

\noindent and

$$
\left(\frac{mn}{n\sigma_{DL,4}^{4}(f,g)+m\sigma_{DL,3}^{2}(f,g)}\right)^{1/2} \biggr(\mathcal{D}^{s}_{KL}(f_n,g)-\mathcal{D}^{s}_{KL}(f,g)\biggr) \rightsquigarrow \mathcal{N}(0, 1).
$$
\end{corollary}

\newpage
\section{PROOFS}\label{sec4}

\bigskip \bigskip
\subsection{The proofs} \label{subsec_mainResults_proofs} $ $\\

\noindent We will begin by the proof of Theorem \ref{thgn}.\\

\noindent \textbf{A - Proof of Theorem \ref{thgn}}.\\

\noindent Suppose that \textbf{Assumptions} \ref{C1} and \ref{C2} are satisfied and $h\in \mathcal{B}_{\infty ,\infty }^{t}\left( \mathbb{R}\right)$.\\

\noindent We have

\begin{eqnarray*}
\int (f_{n}(x)-f(x))h(x)dx&=&(\mathbb{P}_{n,X}(K_{j_{n}}(h))-\mathbb{E}%
_{X}(h) \\
&=&(\mathbb{P}_{n,X}-\mathbb{E}_{X})(K_{j_{n}}(h))+\mathbb{E}_{X}((K_{j_{n}}(h))(X)-h(X)).
\end{eqnarray*}

\noindent It comes that
\begin{equation*}
\mathbb{G}_{n,X}^{w}(h)=\sqrt{n}(\mathbb{P}_{n,X}-\mathbb{E}%
_{X})(K_{j_{n}}(h))\ \ +\ \ \sqrt{n}R_{1,n},
\end{equation*}

\noindent where $R_{1,n}=\mathbb{E}_{X}((K_{j_{n}}(h))(X)-h(X)$.\\

\noindent To complete the proof, we have to show that : (1) $\sqrt{n}(\mathbb{P}_{n,X}-\mathbb{E}_{X})(K_{j_{n}}(h))$ converges in distribution to a centered normal distribution and (2) $\sqrt{n}R_{1,n}$ converges to zero in probability, as $n\rightarrow \infty$. By the way, we will assume that, in the sequel, all the limits as meant as $n\rightarrow \infty$, unless the contrary is specified.\\

\noindent For the first point, we apply the central theorem for independent random variables. We have to check the Lindeberg-Feller-Levy conditions (See \cite{loeve}, Point B, pp. 292). Let us denote $Z_{i,n}=K_{j_{n}}(h)(X_i)$ and 
$\sigma_{i,n}^2=\mathbb{V}ar(Z_{i,n})$, $1\leq i \leq n$ and next $s_{n}^{2}=\sigma_{1,n}^2+\ldots+\sigma_{n,n}^2$, $n \geq 1$. We have to check that
$$
(L1) \  s_{n}^{-1} \max \{\sigma_{i,n}, \ 1\leq i \ \leq n \} \rightarrow 0
$$

\noindent and for any fixed $\varepsilon>0$,

$$
(L2) \ L(n)=:\frac{1}{s_{n}^2} \sum_{i=1}^{n}\int_{(|Y_{i,n}-\mathbb{E}Y_{i,n}|>\varepsilon s_n)}\left\vert Y_{i,n}-\mathbb{E}Y_{i,n}\right\vert^2 d\mathbb{P} \rightarrow 0.
$$

\noindent To prove this, let us begin to see that for any $x\in D$
\begin{eqnarray*}
 \left\vert K_{j_n}(h)(x)-h(x)\right\vert = \int_{D} 2^{j_n} K(2^{j_n}x,2^{j_n}t)(h(t)-h(x))dt.
\end{eqnarray*} 

\noindent By a change of variables and by Assumption \ref{C4}, we have for any $x \in D$,

\begin{eqnarray}
   \left\vert  K_{j_n}h(x)-h(x) \right \vert   & \leq & \int \Phi(u)\left \vert h(x+2^{-j_n}u)-h(x)  \right \vert 1_{(x+2^{-j_n}u \in D)}du.
\end{eqnarray}

\noindent Denote by $C$ a bound of the compact set $\mathcal{K}$ which supports $\Phi$ and $c=\left\Vert\Phi\right\Vert_{\infty} \lambda(\mathcal{K})$. Since $h$ is continuous on the compact set  $D$, it is uniformly continuous and we have
$$
\rho(h,n)=\sup_{(x,t)\in D^2, |t-s|\leq C2^{-j_n}} |f(s)-f(t)| \rightarrow 0,
$$

\noindent which, for all $p\geq  1$, for all $n \geq 1$ and for all $x \in D$, leads to

\begin{eqnarray} \label{F}
   \left\vert  K_{j_n}h(x)-h(x) \right \vert^p f(x)  1_{D}(x) & \leq & c^{p} \rho(h,n)^p f(x) 1_{D}(x) .
\end{eqnarray}

\bigskip \noindent We get that for all $p\geq 1$, for any $1\leq n$. We get some consequences. First, we have

\noindent Then for any $1\leq i \leq n$,
\begin{equation} \label{F2}
 \left\vert  \mathbb{E}Z_{i,n}- \mathbb{E}h(X) \right\vert \leq  \mathbb{E} \left\vert  Z_{i,n}-h(X) \right\vert  \leq  c\rho(h,n) \rightarrow 0
\end{equation}

\noindent Next, $1\leq i \leq n$,

\begin{equation}
\left\vert \left\Vert Z_{i,n} - \mathbb{E}Z_{i,n} \right\Vert_{2} - \left\Vert h(X) - \mathbb{E}Z_{i,n} \right\Vert_{2} \right\vert \leq \left\vert \mathbb{E}h(X) - \mathbb{E}Z_{i,n} \right\vert_{2} \leq c\rho(h,n).
\end{equation}

\bigskip \noindent Hence, the two last formulas yield, 
\begin{equation} \label{unifApprox}
\max_{1\leq i \leq n} |\mathbb{E}Z_{i,n}-\mathbb{E}h(X)| \ \vee \max_{1\leq i \leq n} |\sigma_{i,n}-\mathbb{V}ar(h(X))^{1/2}|  \ \rightarrow 0.
\end{equation}

\noindent Besides, the $c_2$-inequality gives

\begin{eqnarray} \label{c2-01}
\left\vert Z_{i,n}-\mathbb{E}Z_{i,n} \right\vert^2 \leq 2(\left\vert Z_{i,n} \right\vert^2 + \left\vert \mathbb{E}Z_{i,n} \right\vert^2).  
\end{eqnarray}

\bigskip \noindent By applying this $c_2$-inequality to the two terms in the right-hand in Formula (\ref{c2-01}) based on Formulas (\ref{F}) and (\ref{F2}), and by denoting $Z=2(h(X)^2+(\mathbb{E}(h(X))^2$ and
$\delta_n=2c(2+\left\Vert \right\Vert_{\infty})\rho(h,n)$, we have

\begin{eqnarray} \label{c2-02}
\left\vert Z_{i,n}-\mathbb{E}Z_{i,n} \right\vert^2 \leq Z+\delta_n,  
\end{eqnarray}

\bigskip \noindent provided that $n$ is large enough to ensure that $c\rho(h,n)\leq 1$. By the way, we also have
$$
Z+delta_n \leq 6\left\Vert \right\Vert_{\infty}+ \delta_n=\Delta_n \rightarrow 6\left\Vert \right\Vert_{\infty}.
$$

\noindent To prove (L1), put $\alpha(n)=\max \{ |\sigma_{i,n}-\mathbb{V}ar(h(X))^{1/2}| \}$. By (\ref{unifApprox}), we have

$$
\left\vert \frac{s_{n}^2}{n\mathbb{V}ar(h(X))}-1\right\vert \leq max(|(1+\alpha(n))^2-1|,|(1-\alpha(n))^2-1|)\rightarrow 0.
$$

\noindent and then $s_{n}^2 \sim n\mathbb{V}ar(h(X))$. Next

$$
s_{n}^{-1} \max \{\sigma_{i,n}, \ 1\leq i \ \leq n \} \leq \frac{(1+\alpha(n))\mathbb{V}ar(h(X))^{1/2}}{s_n} \sim \frac{(1+\alpha(n))}{\sqrt{n}} \rightarrow 0, 
$$

\noindent which proves (L1). As to (L2), we have uniformly in $1\leq i \leq n$,

\begin{eqnarray*}
L(n)&\leq& \frac{1}{s_{n}^2} \sum_{i=1}^{n}\int_{(Z+\delta_n>\varepsilon^2 s_{n}^2)} \Delta_n d\mathbb{P}\\
&=&\frac{n}{s_{n}^2} \frac{\mathbb{E}Z+\delta_n}{\varepsilon^2s_{n}^2}\\
&\sim & \frac{1}{\mathbb{V}ar(h(X))} \frac{\mathbb{E}Z+\delta_n}{\varepsilon^2s_{n}^2} \rightarrow 0,
\end{eqnarray*}

\bigskip \noindent which proves (L2). Hence the $CLT$ is established and the first point is finished.\\

\noindent As to the second point, we apply Theorem 9.3 in \cite{hard} to have  

\begin{eqnarray*}
\left\vert \mathbb{E}_{X}(K_{j_{n}}h-h)(X)\right\vert &\mathbb{\leq }&\int_E
\left\vert (K_{j_{n}}h)(x)-h(x)\right\vert f(x)dx \\
&\leq &C_{3}\left\Vert (K_{j_{n}}h)-h\right\Vert _{\infty }\left\Vert
f\right\Vert _{\infty } \\
&\leq &\kappa _{2}C_{3}2^{-j_{n}t}.
\end{eqnarray*}

\noindent Therefore, we have
\begin{equation*}
\sqrt{n}R_{1,n}(h)\leq \kappa _{2}C_{3}\sqrt{n}2^{-j_{n}t}=\kappa
_{2}C_{3}n^{(1-2t)/8}=o_{\mathbb{P}}(1),
\end{equation*}

\noindent for any $1/2<t<T$.$\blacksquare$\\

\bigskip \noindent \textbf{B - Proof of Theorem \ref{thJ12}}.\\
\noindent In the proofs, we will systematically use the mean values theorem.  In the multivariate handling, we prefer to use the Taylor-Lagrange-Cauchy as stated
in \cite{valiron}, page 230. The assumptions have already been set up to meet these two rules.
To keep the notation simple, we introduce the two following notations : 

\begin{equation*}
a_{n}=\left\Vert \Delta _{n}f\right \Vert_{\infty} \ \ \text{ and } \ \
b_{n}=\left\Vert \Delta _{n}g\right \Vert_{\infty}.
\end{equation*}

\noindent Recall that 
\begin{equation*}
\mathbb{G}_{n,X}^{w}(h)=\sqrt{n}\int_E\Delta _{n}f(x)h(x)dx\ \ \text{\ \ and \ \ 
} \mathbb{G}_{n,Y}^{w}(h)=\sqrt{n}\int_E\Delta _{n}g(x)h(x)dx,
\end{equation*}

\noindent We start by showing that \ref{thJ12c1} holds.\\

\noindent We have 
\begin{equation*}
\phi (f_{n}(x),g(x))=\phi (f(x)+\Delta _{n}f(x),g(x)).
\end{equation*}

\noindent So by applying the mean value theorem to the function $u_{1}(x)\mapsto \phi(u_{1}(x),g(x))$, we have 
\begin{eqnarray}  \label{fi1}
\phi (f_{n}(x),g(x))&=& \phi (f(x),g(x))\\
&+& \Delta _{n}f(x)\phi_{1}^{(1)}(f(x)+\theta _{1}(x)\Delta _{n}f(x),g(x)) \notag
\end{eqnarray}

\noindent where $\theta _{1}(x)$ is some number lying between $0$ and $1$. In the sequel, any $ \theta_{i}$ satisfies $ \left| \theta_{i} \right|<1$   By applying again the mean values theorem to the function $u_{2}(x)\mapsto
\phi_{1}^{(1)}(u_{2}(x),g(x))$, we have

\begin{eqnarray*}
\Delta _{n}f(x)\phi _{1}^{(1)}(f(x)+\theta _{1}(x)\Delta
_{n}f(x),g(x))&=&\Delta _{n}f(x)\text{ }\phi _{1}^{(1)}(f(x),g(x))\\
&+& \theta_{1}(x)(\Delta _{n}f(x))^{2}\text{ }\phi _{1}^{(2)}(f(x)+\theta
_{2}(x)\Delta _{n}f(x),g(x))\text{ },
\end{eqnarray*}

\noindent where $\theta _{2}(x)$ is some number lying between $0$ and $1$. We can write \eqref{fi1} as 
\begin{eqnarray*}
\phi(f_{n}(x),g(x)) &=& \phi (f(x),g(x))+ \Delta _{n}f(x)\phi_{1}^{(1)}(f(x),g(x))\\
&+&\theta _{1}(x)(\Delta _{n}f(x))^{2}\text{ }\phi_{1}^{(2)}(f(x)\\
&+&\theta _{2}(x)\Delta _{n}f(x),g(x))\text{ }
\end{eqnarray*}

\noindent Now we have
\begin{eqnarray}
\notag J(f_{n},g)-J(f,g)&=&\int_{E} \Delta _{n}f(x)\text{ }\phi_{1}^{(1)}(f(x),g(x))dx \\
\label{fng}
&+&\int_{E} \theta _{1}(x)(\Delta _{n}f(x))^{2}\text{ }\phi _{1}^{(2)}(f(x)+\theta _{2}(x)\Delta _{n}f(x),g(x))\text{ }dx,
\end{eqnarray}

\noindent hence 
\begin{equation*}
|J(f_{n},g)-J(f,g)|\leq a_{n}\int_{E} \left\vert \phi
_{1}^{(1)}(f(x),g(x))\right\vert dx+a_{n}^{2}\int_{E} \left\vert \phi
_{1}^{(2)}(f(x)+\theta _{2}(x)\Delta _{n}f(x),g(x))\text{ }\right\vert dx.
\end{equation*}

\noindent Therefore 
\begin{equation*}
\limsup_{n\rightarrow \infty }\frac{|J(f_{n},g)-J(f,g)|}{a_{n}}\leq A_{1}+
a_{n} \int_{E} \phi _{1}^{(2)}(f(x)+\theta _{2}(x)\Delta _{n}f(x),g(x))\text{
}dx.
\end{equation*}

\noindent Under \textbf{Assumption} \ref{BD}, we know that $A_{1}<\infty$ and that 
\textbf{condition} \eqref{CCS1} is satisfied, that is 

\begin{equation*}
\int_{E} \phi _{1}^{(2)}(f(x)+\theta _{2}(x)\Delta _{n}f(x),g(x))\text{ }%
dx\rightarrow \int_{E} \phi _{1}^{(2)}(f(x),g(x))dx<\infty \ \text{\ \ as \
\ } n\rightarrow \infty.
\end{equation*}
\bigskip
\noindent This proves \eqref{thJ12c1}.\\

\bigskip \noindent Formula \eqref{thJ12c2} is obtained in a similar way. We only need to adapt the result concerning the first coordinate to the second. 
 
 \bigskip \noindent The proof of \eqref{thJ22c1} comes by splitting $\int_{D}\left(\phi(f_{n}(x),g_{m}(x))-\phi(f(x),g(x)) \right) dx$, into the following two terms 

\begin{eqnarray*}
\int_{D} \left(\phi(f_{n}(x),g_{m}(x))-\phi(f(x),g(x)) \right) dx&=& \int_{D}\left(  \phi(f_{n}(x),g_{m}(x))-\phi(f(x),g_{m}(x))\right)  dx\\
&+& \int_{D} \left(\phi(f(x),g_{m}(x))- \phi(f(x),g(x))\right)dx\\
&\equiv & I_{n,1}+ I_{n,2} 
\end{eqnarray*}

\bigskip \noindent We already know how to handle $I_{n,2}$. As to $I_{n,1}$, we may still use the Taylor-Lagrange-Cauchy formula since we have

$$
\left\Vert (f_{n}(x),g_{m}(x))-(f(x),g_{m}(x)\right\Vert_{\infty}=\left\Vert (f_{n}(x)-f(x),0)\right\Vert_{\infty}=a_n\rightarrow 0.
$$

\bigskip \noindent By the Taylor-Lagrange-Cauchy (see \cite{valiron}, page 230), we have

\begin{eqnarray*}
I_{n,1}&=&\int_{D} \Delta f_n(x) \phi(f_{n}(x)+\theta \Delta_{n}f(x),g_{m}(x)) dx\\
&\leq & a_n \int_{D} \phi(f_{n}(x)+\theta \Delta f_n(x),g_{m}(x)) dx\\
&=& a_n (A_2 + o(1)).
\end{eqnarray*}

\noindent From there, the combination of these remarks direct to the result.$\blacksquare$\\

\bigskip \noindent \textbf{C - Proof of Theorem \ref{thJ22}}.\\

\noindent 
\noindent We start by proving \eqref{thJ12n1}. By going back to \eqref{fng}, we have 

\begin{eqnarray*}
\sqrt{n}(J(f_{n},g)-J(f,g)) &=&\sqrt{n}\int_{E} \Delta_{n}f(x)\phi _{1}^{(1)}(f(x),g(x))\text{ }dx \\
&+& \int_{E} \theta _{1}(x)\sqrt{n}\left( \Delta_{n}f(x)\right) ^{2}\text{ }\phi _{1}^{(2)}(f(x)+\theta _{2}(x)\Delta _{n}f(x),g(x))\text{ }dx. \\
&=& \mathbb{G}_{n,X}^w(h_{1}) + \sqrt{n}R_{2,n}
 \end{eqnarray*}
\bigskip \noindent where $R_{2,n}=\int_{E} \theta _{1}(x)\sqrt{n}\left( \Delta_{n}f(x)\right) ^{2}\text{ }\phi _{1}^{(2)}(f(x)+\theta _{2}(x)\Delta _{n}f(x),g(x))\text{ }dx. $ 

\bigskip \noindent Now by Theorem \ref{thgn}, one knows that $\mathbb{G}%
_{n,X}^w(h_{1})\rightsquigarrow \mathcal{N}(0,\mathbb{V}ar(h_{1}(X)) \text{ as }
n\rightarrow \infty$  provided that $h_{1} \in B_{\infty,\infty}^{t}(\mathbb{R)}$. Thus, \eqref{thJ12n1} will be proved if we show that $\sqrt{n}R_{2,n}=0_{%
\mathbb{P}}(1)$. We have  
\begin{equation}
\left\vert \sqrt{n}R_{2,n}\right\vert \leq \sqrt{n}a_{n}^{2}\int_{E} \phi
_{1}^{(2)}(f(x)+\theta _{2}(x)\Delta _{n}f(x),g(x))dx.  \label{r1n}
\end{equation}

\bigskip \noindent Let show that $\sqrt{n}a_{n}^{2}=o_{\mathbb{P}}(1)$. By the Bienaymé-Tchebychev inequality, we have, for any $\epsilon >0$

\begin{equation*}
\mathbb{P}\left( \sqrt{n}a_{n}^{2}>\epsilon \right) =\mathbb{P}\left( a_{n}>%
\frac{\sqrt{\epsilon }}{n^{1/4}}\right) \leq \frac{n^{1/4}}{\sqrt{\epsilon }}%
\mathbb{E}_{X}\left[a_{n}^{2}\right] .
\end{equation*}

\noindent From Theorem 3 in \cite{gine01}, we have 
\begin{eqnarray*}
\left( \mathbb{E}_{X}\left[a_{n}^{2}\right]\right) ^{1/2}&=&O\left( \sqrt{%
\frac{j_{n}2^{j_{n}}}{n}}+2^{-tj_{n}}\right) \\
&=&O\left( \sqrt{\frac{1}{4\log 2} \frac{\log n}{n^{3/4}}} +n^{-t/4} \right)
\end{eqnarray*}

\noindent where we use the fact that $2^{j_{n}}\approx n^{1/4}$. Thus 
\begin{equation*}
\left( \mathbb{P}\left( \sqrt{n}a_{n}^{2}>\epsilon \right)
\right)^{2}=O\left( \sqrt{\frac{1}{4\log 2}\frac{\log n}{n^{1/2}}}+
n^{(1-2t)/8} \right)
\end{equation*}

\noindent Finally $\sqrt{n}a_{n}^{2}=o_{\mathbb{P}}(1)$ since 
\begin{equation*}
\sqrt{\frac{1}{4\log 2}\frac{\log n}{n^{1/2}}}+n^{(1-2t)/8}\rightarrow 0%
\text{ as }n\rightarrow +\infty
\end{equation*}

\noindent for any $t>1/2$. Finally from \eqref{r1n} and using \eqref{CCS1}, we have $\sqrt{n}
R_{2,n}\rightarrow _{\mathbb{P}}0 \text{ as } n\rightarrow +\infty$.\\

\noindent This ends the proof of \eqref{thJ12n1}.

\bigskip \noindent The result \eqref{thJ12n2} is obtained by a symmetry argument by swapping the role of $ f $ and $ g.$

\bigskip \noindent Now, it remains to prove Formula (\ref{thJ22n1}) of the theorem. Let us use bi-variate Taylor-Lagrange-Cauchy formula to get, 
\begin{eqnarray*}  
&&J(f_{n},g_{m})-J(f,g)\\
&=&\int_{E} \Delta_{n}f(x)\phi_{1}^{(1)}(f(x),g(x))dx + \int_{E} \Delta _{m}g(x)\phi _{2}^{(1)}(f(x),g(x))dx \\
&&\frac{1}{2} \int_{D} \biggr(\Delta_{n}f(x)^{2}\phi^{(2)}_{1}+\Delta_{n}f(x) \Delta_{n}g(x)\phi^{(2)}_{1,2}+\Delta_{n}g(x)^{2}\phi^{(2)}_{2}\biggr)\biggr(u_n(x),v_n(y) \biggr)dx.   
\end{eqnarray*}

We have

$$
(u_n(x),v_n(y))=(f(x)+\theta \Delta_{n}f(x), \ g(x)+\theta \Delta_{n}g(x).
$$

\bigskip Thus we get  
\begin{eqnarray*}
J(f_{n},g_{m})-J(f,g) &=&\frac{1}{\sqrt{n}}\mathbb{G}_{n,X}^{w}(h_{1})+\frac{1}{\sqrt{m}}\mathbb{G}_{m,Y}^{w}(h_{2})+R_{n,m},
\end{eqnarray*}

\noindent where $ R_{n,m} $ is given by
$$
\frac{1}{2} \int_{D} \biggr(\Delta_{n}f(x)^{2}\phi^{(2)}_{1}+\Delta_{n}f(x) \Delta_{m}g(x)\phi^{(2)}_{1,2}+\Delta_{m}g(x)^{2}\phi^{(2)}_{2}\biggr)\biggr(u_n(x),v_n(y) \biggr)dx.
$$

\noindent But we have  
\begin{eqnarray*}
\mathbb{G}_{n,X}^{w}(h_{1})&=&N_{n}\left(1
\right)+o_{\mathbb{P}}(1)\\
\mathbb{G}_{m,Y}^{w}(h_{2})&=&N_{n}\left(2\right)+o_{\mathbb{P}}(1),
\end{eqnarray*}

where $ N_{n}\left(i
\right)\sim \mathcal{N}\left(0,\mathbb{V}ar(h_{i}(X)) 
\right), \ \ i=1,2 $ and $N_{n}\left(1
\right)$ and $N_{n}\left(2
\right) $ are independent.

\bigskip \noindent  Using this independence, we have  

\begin{eqnarray*}
\frac{1}{\sqrt{n}} \mathbb{G}_{n,X}^{w}(h_{1})+\frac{1}{\sqrt{m}} \mathbb{G}_{m,Y}^{w}(h_{2})=N\left(0,\frac{
	\mathbb{V}(h_{1}(X))}{n}+\frac{\mathbb{V}(h_{2}(Y))}{m}\right) +o_{\mathbb{P}}\left(\frac{1}{\sqrt{n}}\right)+o_{\mathbb{P}}\left(\frac{1}{\sqrt{m}}\right).
\end{eqnarray*}

\bigskip \noindent Therefore, we have

\begin{eqnarray*}
J(f_{n},g_{m})-J(f,g)=\mathcal{N}\left(0,\frac{\mathbb{V}(h_{1}(X))}{n}+\frac{\mathbb{V}(h_{2}(Y))}{m}\right) +o_{\mathbb{P}}\left(\frac{1}{\sqrt{n}}\right)+o_{\mathbb{P}}\left(\frac{1}{\sqrt{m}}\right)+R_{n,m}.
\end{eqnarray*}

\noindent Hence 
\begin{eqnarray*}
\frac{1}{\sqrt{ \frac{\mathbb{V}(h_{1}(X))}{n}+\frac{\mathbb{V}(h_{2}(Y))}{m}}} \left( J(f_{n},g_{m})-J(f,g)\right)&=&N\left(0,1\right) +o_{\mathbb{P}}\left(\frac{1}{\sqrt{n}} \frac{1}{\sqrt{ \frac{\mathbb{V}(h_{1}(X))}{n}+\frac{\mathbb{V}(h_{2}(Y))}{m}}}\right)\\
&& \ \ \ + \ \ \ o_{\mathbb{P}}\left(\frac{1}{\sqrt{m}}\frac{1}{\sqrt{ \frac{\mathbb{V}(h_{1}(X))}{n}+\frac{\mathbb{V}(h_{2}(Y))}{m}}}\right) \\
&& \ \ \ + \ \ 
\frac{1}{\sqrt{ \frac{\mathbb{V}(h_{1}(X))}{n}+\frac{\mathbb{V}(h_{2}(Y))}{m}}}R_{n,m}.
\end{eqnarray*}

\noindent  That leads to  
\begin{eqnarray*}
\sqrt{\frac{nm}{m\mathbb{V}(h_{1}(X))+n\mathbb{V}(h_{2}(Y))} }\left( J(f_{n},g_{m})-J(f,g)\right)&=&N\left(0,1\right) +o_{\mathbb{P}}(1)\\
&+&\sqrt{\frac{nm}{m\mathbb{V}(h_{1}(X))+n\mathbb{V}(h_{2}(Y))} }R_{n,m},
\end{eqnarray*}

\noindent since $m/(m \mathbb{V}(h_{1}(X))+n \mathbb{V}(h_{2}(Y)))$ and $m/(n \mathbb{V}(h_{1}(X))+n \mathbb{V}(h_{2}(Y)))$ are bounded, and then
\begin{eqnarray*}
o_{\mathbb{P}}\left(\frac{1}{\sqrt{n}} \frac{1}{\sqrt{ \frac{\mathbb{V}(h_{1}(X))}{n}+\frac{\mathbb{V}(h_{2}(Y))}{m}}}\right)&=&o_{\mathbb{P}}\left( \sqrt{\frac{m}{
 m\mathbb{V}(h_{1}(X))+n\mathbb{V}(h_{2}(Y))}}\right)=o_{\mathbb{P}}(1)\\
&& and \\
o_{\mathbb{P}}\left(\frac{1}{\sqrt{m}} \frac{1}{\sqrt{ \frac{\mathbb{V}(h_{1}(X))}{n}+\frac{\mathbb{V}(h_{2}(Y))}{m}}}\right)&=&o_{\mathbb{P}}\left( \sqrt{\frac{n}{
 m\mathbb{V}(h_{1}(X))+n\mathbb{V}(h_{2}(Y))}}\right)=o_{\mathbb{P}}(1).
\end{eqnarray*}

\bigskip \noindent It remains to  prove  that $\left\vert 
\sqrt{\frac{nm}{m\mathbb{V}(h_{1}(X))+n\mathbb{V}(h_{2}(Y))} }R_{n,m}\right\vert =o_{\mathbb{P}}(1).$ But we have by the continuity assumptions on $\phi$ and on its partial derivatives and by the uniform of $\Delta_{n}f(x)$ and $\Delta_{n}g(x)$ to zero, that

\begin{eqnarray*}
&&\left\vert \sqrt{\frac{nm}{m\mathbb{V}(h_{1}(X))+n\mathbb{V}(h_{2}(Y))} }R_{n,m}\right\vert \leq\\
&&\frac{1}{2} \left(\sqrt{n}a_{n}^{2} (\int_{D} \phi^{(2)}_{1}(f(x),g(x))dx +o(1))  \right) \left( \sqrt{\frac{m}{ m\mathbb{V}(h_{1}(X))+n\mathbb{V}(h_{2}(Y))}}\right)\\
&+&\frac{1}{2} \left(\sqrt{m}b_{m}^{2} (\int_{D} \phi^{(2)}_{2}(f(x),g(x))dx +o(1))  \right) \left( \sqrt{\frac{n}{ m\mathbb{V}(h_{1}(X))+n\mathbb{V}(h_{2}(Y))}}\right)\\
&+&\frac{1}{2} \left(\sqrt{n}a_{m}b_{m} (\int_{D} \phi^{(2)}_{2}(f(x),g(x))dx +o(1))  \right) \left( \sqrt{\frac{n}{ m\mathbb{V}(h_{1}(X))+n\mathbb{V}(h_{2}(Y))}}\right)
\end{eqnarray*}

\bigskip \noindent As previously, we have $\sqrt{n}a_{n}^{2}=o_{\mathbb{P}}(1)$, $\sqrt{m}b_{m}^{2}=o_{\mathbb{P}}(1)$ and $\sqrt{n}a_{m}b_{m}=o_{\mathbb{P}}(1)$.

\bigskip \noindent From there, the conclusion is immediate.$\blacksquare$\\

\newpage
\section{Annexe}\label{annexe}

\noindent Here, we address the applicability our results on usual distribution functions. We have seen that we need to avoid infinite and null values. For example, integrals in the Renyi's of the Tsallis family, we may encounter such problems as signaled in the first pages of this paper. To avoid them, we already suggested to used a modification of the considered divergence measure in the following way.\\

\noindent First of all, it does not make sense to compare two distributions of different supports. Comparing a \textit{pdf} with support 
$\mathbb{R}$, like the Gaussian one, with another with support $[0,1]$, like the standard uniform one, is meaningless. So, we suppose that the \textit{pdf}'s we are comparing have the same support $D$.\\

\noindent Next, for each $\varepsilon>0$, we find a domain $D_{\varepsilon}$ included in the common support $D$ of $f$ and $g$ such that

\begin{equation}
\int_{D_{\varepsilon}} f(x) dx \geq 1-\varepsilon  \text{ and  }  \int_{D_{\varepsilon}} g(x) dx \geq 1-\varepsilon. \label{DomaineEpsi}
\end{equation} 
 
\noindent and there exist two finite numbers $\kappa _{1}>0$ and $\kappa _{2}>0$, such that we have

\begin{equation}
\kappa _{1}\leq f 1_{D_{\varepsilon}}, g 1_{D_{\varepsilon}}\leq \kappa _{2}.  \label{BDEN}
\end{equation} 

\bigskip \noindent Besides, we choose the $D_{\epsilon}$'s increasing to $D$ as $\epsilon$ decreases to zero. We define the modified divergence measure

\begin{equation}
\mathcal{D}^{(\varepsilon)}(f,g)=\mathcal{D}(f 1_{D_{\varepsilon}},g 1_{D_{\varepsilon}}). \label{modifDiv}
\end{equation}

\noindent We may denote 
$$
f_{\varepsilon}=f 1_{D_{\varepsilon}} \text{ and } g_{\varepsilon}=g 1_{D_{\varepsilon}}. 
$$

\
\noindent Based on the remarks that the $D_{\epsilon}$'s increasing to $D$ as $\epsilon$ decreases to zero and that the equality between $f$ and $g$ implies that of $f_{\varepsilon}$ and $g_{\varepsilon}$, we recommend to replace the exact test of $f=g$ by the approximated test $f_{\varepsilon}=g_{\varepsilon}$, for $\varepsilon$ as small as possible.\\

\noindent So each application should begin by a quick look at the domain $D$ of the two \textit{pdf} and the founding of the appropriate sub-domain 
$D_{\varepsilon}$ on which are applied the tests.\\

\noindent Assumption (\ref{BDEN}) also ensures that the \textit{pdf}'s $f_{\varepsilon}$ and $g_{\varepsilon}$ lie in $\mathcal{B}^{t}_{\infty \infty}$ for almost all the usual laws. Actually, according to \cite{hard}, page 104, we have that $f\in \mathcal{B}^{t}_{\infty \infty}$, for some $t>0$, if and only if
$$
\sup_{x \in \mathbb{R}} |f(x)| + \sup_{x \in \mathbb{R}} \sup_{h\neq 0} \frac{f^{[t]}(x+h)-2f^{[t]}(x)+f^{[t]}(x-h)}{|h|^{t-[t]}},
$$

\noindent where $[t]$ stands for the integer part of the real number $t$, that is the greatest integer less or equal to $f$ and $f^p$ denotes the $p$-th derivative function of $f$.\\

\noindent Whenever the functions $f_{\varepsilon}$ and $g_{\varepsilon}$ have $([t]+1)$-th derivatives bounded and not vanishing on $D_{\varepsilon}$, they will belong to $f\in \mathcal{B}^{t}_{\infty \infty}$. Assumption (\ref{BDEN}) has been set on purpose for this. Once this is obtained, all the functions that are required to lie on $\mathcal{B}^{t}_{\infty \infty}$ for the validity of the results, effectively are in that space. All examples we will use in this sections satisfy these conditions, including the following random variables to cite a few : Gaussian, Gamma, Hyperbolic, etc.\\

\bigskip \noindent \textbf{Acknowledgment}
The fourth (1 \& 2 \& 3) author acknowledges support from the World Bank Excellence Center (CEA-MITIC) that is continuously funding his research activities from starting 2014.\\

\end{document}